\documentclass[12pt]{amsart}
\usepackage{graphicx}
\usepackage{caption}
\usepackage{subcaption}
\usepackage{setspace}
\usepackage{empheq}
\usepackage{cite}
\usepackage{amsmath}
\usepackage{mathtools}
\usepackage{textcomp}
\usepackage{bm}
\usepackage{amssymb}

\usepackage{color}
\definecolor{MyLinkColor}{rgb}{0,0,0.4}

\newcommand{\e}{\varepsilon}

\newcommand{\f}{\frac}
    \DeclareMathOperator{\csch}{csch}

\newtheorem{thm}{Theorem}[section]

\theoremstyle{remark}
\newtheorem{rem}[thm]{Remark}

\numberwithin{equation}{section}

\title[Interfacial internal waves over variable bottom]{Hamiltonian approach to modelling interfacial internal waves over variable bottom}
\author[R. I. Ivanov ]{Rossen I. Ivanov }
\address{School of Mathematical Sciences, Technological University Dublin, City Campus, Grangegorman Lower, Dublin, D07 ADY7 Ireland}
\email{rossen.ivanov@tudublin.ie}
\author[C. I. Martin]{Calin I. Martin }
\address{Faculty of Mathematics, University of Vienna, Oskar-Morgenstern-Platz 1, 1090 Vienna, Austria}
\email{calin.martin@univie.ac.at}
\author[M. D. Todorov]{Michail D. Todorov }
\address{Department of Differential Equations, Faculty of Applied Mathematics and Informatics, Technical University of Sofia, 8 Kliment Ohridski Boulevard, 1000 Sofia, Bulgaria}
\email{mtod@tu-sofia.bg}

\begin{document}
 \begin{abstract}
We study the effects of an uneven bottom on the internal wave propagation in the presence of stratification and underlying non-uniform currents. Thus, the presented models incorporate vorticity (wave-current interactions), geophysical effects (Coriolis force) and a variable bathymetry. An example of the physical situation described above is well illustrated by the equatorial internal waves in the presence of the Equatorial Undercurrent (EUC). We find that the interface (physically coinciding with the thermocline and the pycnocline) satisfies in the long wave approximation a KdV-mKdV type equation with variable coefficients. The soliton propagation over variable depth leads to effects such as soliton fission, which is analysed and studied numerically as well.
\end{abstract}

\maketitle

\section{Introduction}

It is well known that ocean wave dynamics displays very complex features, partly due to the effects of the bottom topography which quite often deviates from the convenient scenario of a flat bed. Waves over variable bottom are an active area of research, and various scales, geometries and approximations have been examined. For a survey of results we refer the reader, for example, to the review by Kirby \cite{Kir}, the book by Dingemans \cite{Din} and the references therein. The best known nonlinear wave models have been extended for fluids with uneven bottom as well: we recall here the KdV equation for long waves, for example, which has been generalised and studied thoroughly by Johnson \cite{Johnson71}, and the NLS equation for modulated waves, treated by Djordjevi\'c and Redekopp \cite{DR}.

The appreciable body of studies handling free surface and/or internal water wave propagation over variable depth covers situations like waves in channels (Rosales and Papanicolaou) \cite{RoPa}, non-hydrostatic topographic effects \cite{CHL},  rapidly varying topographies \cite{N,LN}, surface waves over internal waves \cite{CGS} and currents \cite{KD}, and even tsunami generation \cite{DD}. Higher order nonlinearities and dispersion as well as intermediate long wave propagation regimes have been examined by Choi and Camassa \cite{Choi,Choi2,Choi3}. Internal waves over variable bottom have been studied extensively as well \cite{Choi,Mad,Grim,Grim10,HM,DR,DR2,RW}. 

While most of the studies of internal waves involve irrotational flow, shear background currents have been included as well \cite{RW,CIMT,CompelliIvanov1,CJ15,CI-19,CIM-16,Iv17,JCI}. However, the combined effects of a variable bottom topography, sheared currents and stratification on the arising internal waves are rather less investigated. We attempt to bring our contribution toward filling this gap by a derivation of a model equation (with variable coefficients) of  KdV type which describes the interface in a flow with a variable depth and a flat surface in the presence of currents, density stratification and geophysical effects. A significant part in our endeavour is played by a variational approach based on the Zakharov's Hamiltonian formulation \cite{Z68}, the subsequent developments such as \cite{BB1,BB2,BO} as well as other irrotational scenarios, like surface and internal waves \cite{CGK} or variable bottoms \cite{CGNS,CIT}. We advance here another feature of ocean dynamics: the presence of (non-uniform) underlying currents modeled by a specific choice of vorticity function.
 
The Hamiltonian formulation in our approach makes an extensive use of the Dirichlet-Neumann operators (DNO) \cite{CG,CGK,CGNS} and provides a convenient setting for incorporating a series of features like interacting fluid layers, topography effects, stratification and underlying currents. In the presented study we illustrate the derivation and application of the Hamiltonian framework based on DNO for the case of uneven bottom. The illustrative example concerns internal wave propagation in the equatorial region in the presence of the Equatorial Undercurrent. The equatorial internal waves are special in a sense due to the effect of the Coriolis force. This effect keeps the waves propagating along the Equator like in a wave guide. Some further details could be found for example in \cite{CJ15,CI-19}. We would like to note also the recent study by Guyenne \cite{Guy} who proposed a numerical model for nonlinear surface waves in the presence of a vertically sheared current utilizing a Hamiltonian formulation combined with a series expansion of the DNO.

The rest of the paper is organised as follows. In Section 2 we formulate the model from the governing equations of the fluid mechanics. In Section 3 we describe the Hamiltonian representation of the evolution equations and then in Section 4 we present a long wave approximation which reduces to a KdV equation with variable coefficients which represents a generalisation of the flat bottom scenario. The effects on the solitons such as fission due to the variable depth are studied in Section 5. A special case with small or vanishing coefficient of the quadratic nonlinearity, taking into account the cubic nonlinearity, is analysed in Section 6. The analogues of the basic conserved quantities are given in Section 7. The Dirichlet-Neumann operators for the lower layer which is bounded by an uneven bottom is derived in the Appendix A. The numerical scheme for the finite-difference implementation of KdV-type equation with variable coefficients is outlined in Appendix B.

\section{Equations of motion for an internal wave}
\begin{figure}
\begin{center}
\fbox{\includegraphics[totalheight=0.30\textheight]{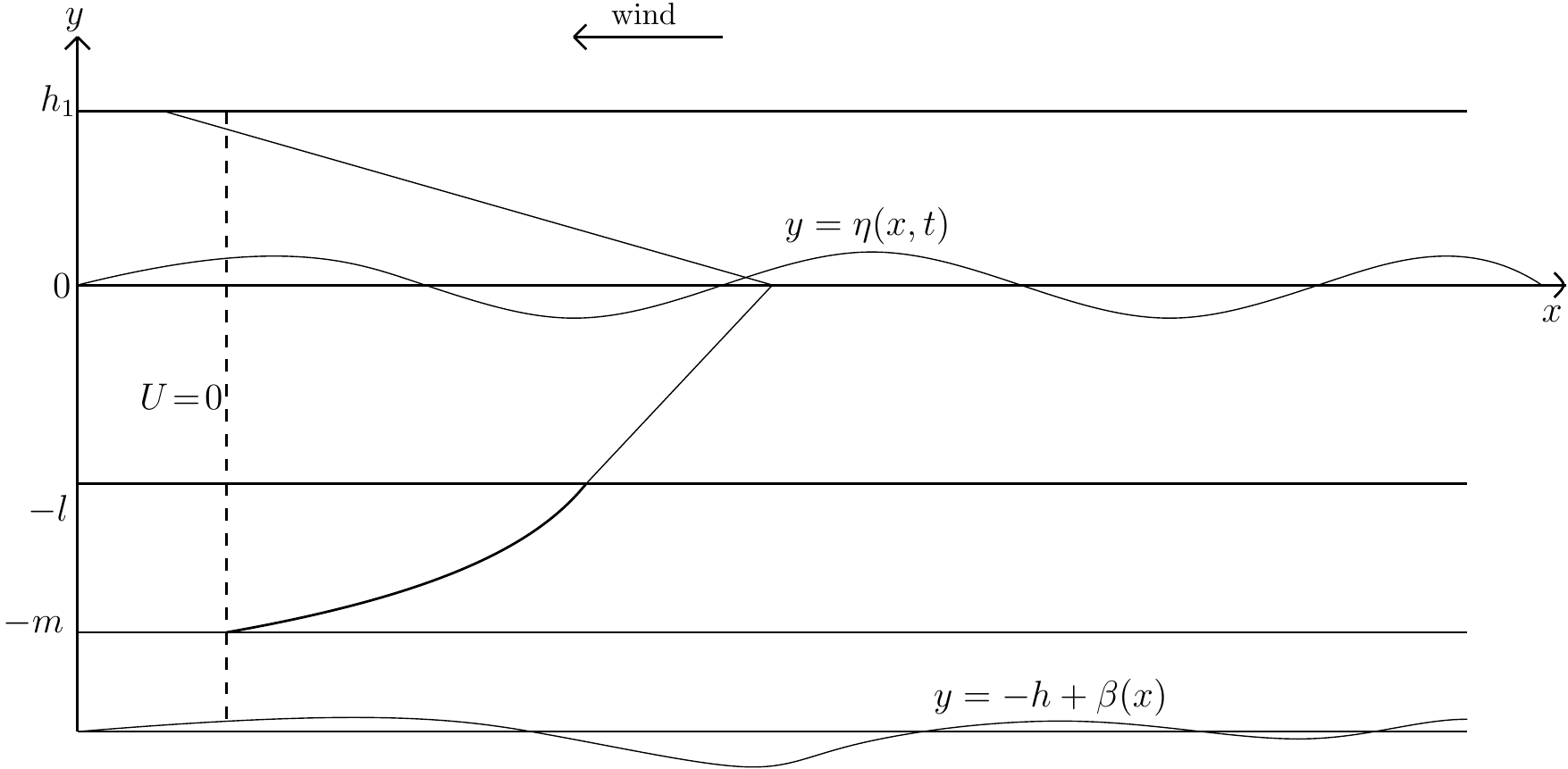}}
\caption{The system under study involving the two domains of different density \eqref{topl}, \eqref{botl} and the four layers of different vorticity \eqref{U5}. The flat surface is $y=h_1,$ the elevation of the internal wave corresponds to the curve $y=\eta(x,t)$ and the variable bottom corresponds to $y=-h+\beta(x).$ The current profile \eqref{U5} (for the special case of undisturbed interface $\eta=0$) is given as well, the magnitude $\bm{U}$ is oriented along the horizontal axis, and the vertical dashed line corresponds to $\bm{U}=0.$ }
\label{fig:thesisfigure_system}
\end{center}
\end{figure}

We consider a two-dimensional water flow, moving under the influence of gravity, such that the $x$-axis is oriented along the horizontal direction and the $y$-axis is pointing vertically upwards. The time variable will be denoted with $t$.
The flow is composed of two domains, $\Omega$ and $\Omega_1$, consisting of water with different constant densities and separated by a common interface, denoted $y=\eta(x,t)$, which represents an interfacial internal wave. More specifically,
we assume that, adjacent to the bed, the lower water domain is given as
\begin{equation} \label{botl} \Omega(\eta, \beta):=\{(x,y,t):x\in\mathbb{R},\,\,t\in\mathbb{R},\,\,
\mathcal{B}(x):=-h+\beta(x)<y<\eta(x,t)\},\end{equation} being situated below the near-surface region
\begin{equation}\label{topl}\Omega_1(\eta):=\{(x,y,t):x\in\mathbb{R},\,\,  t\in\mathbb{R},\,\,\eta(x,t)<y<h_1\},\end{equation}
where $\beta(x)$ is some given function, indicative of the unevenness of the bottom, 
while $h$ and $h_1$ are positive constants, such that $y=h_1$ is the flat surface, and $y=-h$ is the average depth of the seafloor, cf. Figure \ref{fig:thesisfigure_system}.
Moreover, we assume that the internal wave has the properties
 $$\lim_{|x|\to\infty}\eta(x,t)=0, \qquad \int_{\mathbb{R}}\eta(x,t)dx=0,$$  the last one related to the fact that the average depth of the interface is at $y=0.$ \footnote{Since the average depth involves division by the length of the interval (which is infinity, the interval is the whole real axis) the average depth will be the same if  $\int_{\mathbb{R}}\eta(x,t)dx$ is a finite constant.}
\noindent Taking into account the Equator's peculiar feature of behaving like a wave-guide we will look at two-dimensional  inviscid and incompressible fluid motion confined near the Equator by the action of the Coriolis forces.

We would like to point out that most of the physical variables that we employ here (like the density, the generalised velocity potentials or the components of the velocity field) display discontinuities across the interface $y=\eta(x,t)$ that separates the two fluid regions. To make the reader observant of this aspect, we use the index $1$ as a label for the upper layer. Whenever we refer to the overall physical variable without specification of the layer, we shall use bold face symbol.

Therefore, denoting with $(\bm{u}(x,y,t),\bm{v}(x,y,t))$ the velocity field, the equations of motion are Euler's equations
\begin{equation}
 \left\{\begin{array}{lcl}
        \bm{ u}_t+\bm{u}\bm{u}_x+\bm{v}\bm{u}_y+2\omega \bm{v} &=& -\bm{\frac{1}{\rho}}P_x,\\
         \bm{v}_t+\bm{u}\bm{v}_x+\bm{v}\bm{v}_y -2\omega \bm{u} &=& -\bm{\frac{1}{\rho}}P_y-g,
        \end{array}\right.
\end{equation}
where $P=P(x,y,t)$ denotes the pressure, $\omega$ is the rotational speed of Earth, $g$ is the gravitational acceleration and $\bm{\rho}$ denotes the density of the fluid which is assumed to be piece-wise constant being distributed as 
\begin{equation}
\bm{\rho}(x,y)=
\left\{\begin{array}{c}\rho_1(x,y)\,\,{\rm for}\,\,(x,y)\in \Omega_1,\\ \rho(x,y)\,\,{\rm for}\,\,(x,y)\in \Omega,\end{array}\right.
\end{equation}
with the understanding that we look at a stable stratification, that is $\rho>\rho_1$.
Throughout the paper we will consider rotational water flows, and moreover, we presuppose that the vorticity is given as
$$\bm{\gamma}:=\bm{u}_y-\bm{v}_x=\bm{U}'(y),$$
where $\bm{U}(y)$ represents the background shear current and has the following profile: the current is piece-wise linear with respect to $y$ with the exception of the layer near the bottom where it decays to zero. More specifically, we introduce five layers as follows : 
\begin{equation} \label{U5}
\bm{U}(y)=\left\{\begin{array}{lll}\gamma_1 y+\kappa, &\rm{when} & \eta(x,t) \le y \le h_1,\\ 
\gamma y + \kappa, & \rm{when} & -l\le y \le \eta(x,t) \\
U(y)&\rm{when} & -m  \le y \le -l \\
0 &\rm{when} & -h +\beta(x) \le y \le -m.\end{array}\right.
\end{equation} Here  $\gamma$ and $\gamma_1$ represent constant vorticities in the corresponding sub-domains, $\kappa$ is a constant component of the current. The choice of this current is made to model the Equatorial undercurrent which is formed by the winds blowing to the west \cite{CI-19,CIM-16,CompelliIvanov1} so that the current on the surface is negative. We note that the current is not continuous at $y=\eta$, and has a jump when $\eta\ne 0$ (continuous is only the normal component of the velocity field). Between the other layers $U(y)$ is  a continuous function such that $U(-m)=0,$ $U(-l)=-l\gamma+\kappa.$  On Fig. \ref{fig:thesisfigure_system} the current is illustratively shown for the situation of an  undisturbed fluid when $\eta\equiv 0.$
The layer between depths $y=-m$ and $y=-l$ is with prescribed current shape $U(y)$ which could be chosen, for example, as necessary to match the field data. It could be shown that the energy per unit length of this layer is constant and thus the layer does not contribute to the equations on the interface (the pycnocline) \cite{CompelliIvanov1}. However, the reconstruction of the velocity field in the body of the fluid would, of course, depend on $U(y).$  The current gradually weakens towards the bottom layer, where it is zero, and there the motion is irrotational.

With respect to the described stratification we have the following notations for the velocity field
\begin{equation}
\bm{u}(x,y,t):=
\left\{\begin{array}{ccc}  u(x,y,t), &{\rm in} & \Omega,\\ u_1(x,y,t), &{\rm in} &\Omega_1,\end{array}\right.
\end{equation}
and
\begin{equation}
\bm{v}(x,y,t):=
\left\{\begin{array}{ccc} v(x,y,t), &{\rm in} & \Omega,\\ v_1(x,y,t), &{\rm in} &\Omega_1.\end{array}\right.
\end{equation}
In addition, we have the equation of mass conservation for incompressible fluid
\begin{equation}\label{masscons}
\bm{ u}_x+\bm{v}_y=0\,\,\textrm{in}\,\,\Omega\cup\Omega_1.
\end{equation}
Complementing the equations of motion are the boundary conditions, of which
the dynamic boundary condition
\begin{equation}\label{atm}
 P=P_{atm}\,\,{\textrm on}\,\, y=h_1,   
\end{equation}
(with $P_{atm}$ being the constant atmospheric pressure) decouples the motion of the water from that of the air.
In addition, the impermeability of the surface boundary, of the interface and of the bottom,  
leads to the kinematic boundary conditions. On the surface and the interface respectively they are
\begin{equation}
 \label{kin_fs}
 v_1=0\,\,{\rm on}\,\,y=h_1,
\end{equation}
\begin{equation}
 \label{kin_int}
 \begin{array}{lll}
v_1=\eta_t +u_1 \eta_x & {\rm on} & y=\eta(x,t),\\
v=\eta_t+u\eta_x & {\rm on} & y=\eta(x,t),
\end{array}
\end{equation}
while on the bottom we have the condition
\begin{equation}
\label{KBC_VarBott}
\varphi_y (x,\mathcal{B}(x))=\mathcal{B}_x \varphi_x (x,\mathcal{B}(x)),\,\,{\rm for}\,\,{\rm all}\,\,x\in\mathbb{R},
\end{equation}
stating that the normal component of the velocity field vanishes.

To ease the notation, in the further developments of the paper, we introduce sub-indices $b$ and $s$ for the evaluations of physical quantities at the bottom $y=\mathcal{B}(x)$ and at the common interface $y=\eta(x,t)$, respectively. In order to describe propagation of solitary waves
 we make the assumption that all considered functions $\eta(x,t)$, $\varphi(x,y,t)$, $\varphi_1(x,y,t)$,  $\beta(x)$ are in the Schwartz class with respect to the $x$ variable, that is declining fast enough when $x\to \pm \infty$ (for all values of the other variables), which we denote by $\eta(x, \cdot)\in \mathcal{S} (\mathbb{R}),$ etc.

The equation of mass conservation \eqref{masscons} ensures the existence of a stream function
$$\bm{\psi}(t,x,y)=\left\{\begin{array}{ccc}\psi(t,x,y)& {\rm in}&\Omega,\\ \psi_1(t,x,y) & {\rm in} & \Omega_1,\end{array}\right.$$
determined up to an additive term that depends only on time, by
\begin{equation}\label{stream_func}
\left\{
\begin{array}{lll}
 u=\psi_y, & v=-\psi_x, & {\rm in} \quad \Omega,\\
u_1=\psi_{1,y}, & v_1=-\psi_{1,x}, & {\rm in}\quad\Omega_1
\end{array}\right.
\end{equation}
We will impose the condition that the stream function is continuous on the interface $y=\eta(x,t)$, condition which can be expressed as
\begin{equation}
 \psi(t,x,\eta(x,t))=\psi_1(t,x,\eta(x,t)).
\end{equation}
The latter condition implies that the normal velocity components equal across the interface $y=\eta(x,t)$.

We introduce now the (generalized) velocity potential $$\bm{\varphi}=\left\{\begin{array}{ccc} \varphi, &{\rm in}& \Omega,\\ \varphi_1, &{\rm in} & \Omega_1,\end{array}\right.$$
by means of
\begin{equation}\label{vel_pot}
   \bm{u}=\bm{\varphi}_x+ \bm{U}(y) , \qquad \bm{v}=\bm{\varphi}_y,
\end{equation}
The kinematic boundary conditions \eqref{kin_fs} and \eqref{kin_int} can now be written as
\begin{equation}\label{kin_surface}
 (\varphi_y)_b=\mathcal{B}_x(\varphi_x)_b, \quad  (\varphi_{1,y})_{y=h_1}=0
\end{equation}
and respectively, as
\begin{equation}\label{kin_interface}
 \begin{array}{c}
   \eta_t=( \varphi_{1,y})_{s}-\eta_x [(\varphi_{1,x})_{s}+\gamma_1\eta+\kappa],\\
 \eta_t=(\varphi_y)_{s}- \eta_x [(\varphi_x)_{s}+\gamma\eta+\kappa].
  \end{array}
\end{equation}
The following notation will be used later on in the paper. Namely, we set
\begin{equation}\label{Phi_notation}
\begin{array}{l}
 \Phi(x,t)=\varphi(x,\eta(x,t),t),\\
 \Phi_1(x,t)=\varphi_1(x,\eta(x,t),t),\\
 \Phi_b(x,t)=\varphi(x,\mathcal{B}(x),t).
 \end{array}
\end{equation}

Euler's equations can be expressed by means of the stream function and of the generalized velocity potential as
\begin{equation}
\nabla\left[\bm{\varphi}_t+\frac{1}{2}|\nabla\bm{\psi}|^2+\frac{P}{\rho}-(\bm{\gamma}+2\omega)\bm{\psi}+gy\right]=0.
\end{equation}
We have therefore
$$\varphi_{1,t}+\frac{1}{2}|\nabla\psi_1|^2 -(\gamma_1+2\omega)\psi_1 +\frac{P_1}{\rho_1}+gy=f_1(t)\quad{\rm in}\quad \Omega_1,$$
$$\varphi_{t}+\frac{1}{2}|\nabla\psi|^2 -(\gamma+2\omega)\psi +\frac{P}{\rho}+gy=f(t)\quad{\rm for}\quad -l\le y \le \eta(x,t).$$
From the condition $P=P_{atm}$ on the top surface and availing also of the Schwartz property of the stream function $\psi_1$ and of the velocity potential $\varphi_1$ we infer that $P_{atm}/\rho_1+gh_1=f_1(t)$ for all $t$.
 Moreover, utilizing the continuity of the pressure at the interface $y=\eta(x,t)$ and making the choice $f(t):=\rho_1 f_1(t)/\rho$  we obtain the Bernoulli type equation
\begin{equation}
 \label{Bint}
 \rho\left[(\varphi_t)_s+\frac{|\nabla\psi|_s^2}{2}-(\gamma+2\omega)\chi+g\eta\right]=\rho_1\left[(\varphi_{1,t})_s+\frac{|\nabla\psi_1|_s^2}{2}-(\gamma_1+2\omega)\chi+g\eta\right]
\end{equation}
where $\chi:=\psi(t,x,\eta(x))=\psi_1(t,x,\eta(x))$ is the stream function evaluated at the interface.

\section{The Hamiltonian functional and the Hamiltonian formulation}
We start this section by indicating the most physical choice for the Hamiltonian functional which is the total energy of the flow. This functional is then written in terms of the canonical variables: the interface defining function $\eta$ and a suitable combination of the velocity potentials $\varphi$ and $\varphi_1$ evaluated on the interface. We then introduce an almost-Hamiltonian formulation of the considered wave-current system that becomes Hamiltonian in the absence of sheared underlying currents (that is, for $\gamma=\gamma_1=0$).
Lastly, it will be proven that an appropriate change of variables renders the almost-Hamiltonian formulation into a \emph{bona fide} Hamiltonian one.
\subsection{The Hamiltonian}
The outset of this section's endeavour is the evaluation of the kinetic and of the potential energy for each domain. For the lower domain the potential energy is 
\begin{equation}
V_l[\eta]= \frac{\rho g}{2} \int_{\Omega} y dy dx = \frac{\rho g}{2}\int_{\mathbb{R}} (\eta^2-\mathcal{B}^2) dx.  \nonumber \\
\end{equation}
Observing that 
\begin{equation}
u^2+v^2=(\varphi_x + \bm{U}(y))^2 + \varphi_y^2=\textrm{div} (\varphi \nabla \varphi) + 2 \varphi_x \bm{U(y)}+ (\bm{U(y)})^2
\end{equation}
we use Green's Theorem (Divergence Theorem)  and find that the kinetic energy of the lower layer can be calculated as 
\begin{align}
\label{K-L}
K_l[\eta, \Phi]&=\frac{\rho}{2}\int_{\Omega} \mathrm{div}(\varphi \nabla \varphi) dy dx +\rho\int_{\Omega} \bm{U}(y) \varphi_x dy dx +\frac{\rho}{2}\int_{\Omega} \bm{U}^2(y)dy dx  \nonumber \\
&= \frac{\rho}{2}\int_{\mathbb{R}} (\varphi \nabla \varphi)_s \cdot {\bf n}_s \sqrt{1+\eta_x^2}dx +\frac{\rho}{2}\int_{\mathbb{R}} (\varphi \nabla \varphi)_b \cdot {\bf n}_b \sqrt{1+\mathcal{B}_x^2}dx  \nonumber \\
&- \rho \int_{\mathbb{R}} (\gamma \eta + \kappa)\Phi(x,t) \eta_x dx +\frac{\rho}{2}\int_{\mathbb{R}}\int_{-m}^{-l} {U}^2(y)dy dx \nonumber \\
& + \frac{\rho}{6 \gamma} \int_{\mathbb{R}}[(\gamma \eta + \kappa)^3-(-\gamma l + \kappa)^3] dx,
\end{align}

\noindent where  ${\bf n}_s = (-\eta_x, 1)/\sqrt{1+\eta_x^2}$ is the outward-pointing unit normal vector (with respect to $\Omega$) to the wave surface, 
while ${\bf n}_b=(\mathcal{B}_x, -1)/\sqrt{1+\mathcal{B}_x^2}$ is
the outward-pointing unit normal vector on the bottom.
Since $$\sqrt{1+\mathcal{B}_x^2}\, \, {\bf n}_b \cdot (\nabla \varphi)_b =(\mathcal{B}_x, -1) \cdot \left(( \varphi_x)_b, (\varphi_y)_b\right)=(\mathcal{B}_x( \varphi_x)_b- (\varphi_y)_b)=0  $$ we have that no bottom-related terms are present. Hence
\begin{align}
\label{K-L1}
K_l[\eta, \Phi]&= \frac{\rho}{2}\int_{\mathbb{R}} \Phi (\nabla \varphi)_s \cdot {\bf n}_s \sqrt{1+\eta_x^2}dx   \nonumber \\
&- \rho \int_{\mathbb{R}} (\gamma \eta + \kappa)\Phi(x,t) \eta_x dx +\frac{\rho}{2}\int_{\mathbb{R}}\int_{-m}^{-l} {U}^2(y)dy dx \nonumber \\
& + \frac{\rho}{6 \gamma} \int_{\mathbb{R}}[(\gamma \eta + \kappa)^3-(-\gamma l + \kappa)^3] dx.
\end{align}

Let us introduce the Dirichlet-Neumann operators $G_{ij}(\beta,\eta)$ given by
\begin{equation}
\begin{pmatrix}
G_{11} & G_{12}\\
G_{21} & G_{22}
\end{pmatrix} \begin{pmatrix}\Phi \\ \Phi_b \end{pmatrix}=
\begin{pmatrix}
(\nabla\varphi)_s\cdot {\bf n}_s\sqrt{1+\eta_x^2}\\
(\nabla\varphi)_b\cdot {\bf n}_b\sqrt{1+\mathcal{B}_x^2}
\end{pmatrix}.
\end{equation}
Therefore \begin{align}
& G_{11}\Phi + G_{12}\Phi_b=(\nabla\varphi)_s\cdot {\bf n}_s\sqrt{1+\eta_x^2}, \nonumber \\
& G_{21}\Phi + G_{22}\Phi_b =0, \qquad  \Phi_b = -G_{22}^{-1}G_{21}\Phi,  \end{align} and
\begin{equation}
(\nabla\varphi)_s\cdot {\bf n}_s\sqrt{1+\eta_x^2} = \left(G_{11}-G_{12}G_{22}^{-1}G_{21} \right)\Phi.
\end{equation}
The expression \eqref{K-L1} becomes
\begin{align}
\label{K-L2}
K_l[\eta, \Phi]&= \frac{\rho}{2} \int_{\mathbb{R}} \Phi \left(G_{11}-G_{12}G_{22}^{-1}G_{21} \right)\Phi dx   \nonumber \\
&- \rho \int_{\mathbb{R}} (\gamma \eta + \kappa)\Phi(x,t) \eta_x dx +\frac{\rho}{2}\int_{\mathbb{R}}\int_{-m}^{-l} {U}^2(y)dy dx \nonumber \\
& + \frac{\rho}{6 \gamma} \int_{\mathbb{R}}[(\gamma \eta + \kappa)^3-(-\gamma l + \kappa)^3] dx.
\end{align}
Let us denote $G(\eta, \beta):=G_{11}-G_{12}G_{22}^{-1}G_{21}.$ This operator depends on the bottom variations through $\beta(x).$ The details of the computation of these Dirichlet-Neumann operators is given in the Appendix.

For the upper layer, similarly,
\begin{align}
\label{KP-U}
K_u[\eta, \Phi_1]&= \frac{\rho_1}{2} \int_{\mathbb{R}} \Phi_1 G_1 (\eta)\Phi_1 dx   \nonumber \\
&+ \rho_1 \int_{\mathbb{R}} (\gamma_1 \eta + \kappa)\Phi_1(x,t) \eta_x dx  \nonumber \\
& + \frac{\rho_1}{6 \gamma_1} \int_{\mathbb{R}}[(\gamma_1 h_1 + \kappa)^3-(\gamma_1 \eta + \kappa)^3] dx, \nonumber\\
V_u[\eta]&= \frac{\rho_1 g}{2} \int_{\Omega_1} y dy dx = \frac{\rho_1 g}{2}\int_{\mathbb{R}} (h_1^2-\eta^2) dx 
\end{align}
where $G_1(\eta)$ is defined as 
\begin{equation} \label{G1def}
    G_1(\eta) \Phi_1=-(\nabla\varphi_1)_s\cdot {\bf n}_s\sqrt{1+\eta_x^2}=-(\varphi_{1,y})_s +(\varphi_{1,x})_s\eta_x.
\end{equation}The minus sign is because the outward normal for the domain $\Omega_1$ is $- {\bf n}_s.$
Recall that 
 \begin{equation} \label{Gdef}
    G(\eta,\beta) \Phi=(\nabla\varphi)_s\cdot {\bf n}_s\sqrt{1+\eta_x^2}=(\varphi_{y})_s -(\varphi_{x})_s\eta_x.
\end{equation}

Some of the integrals above are not convergent due to the constant densities at infinity. However, the Hamiltonian is the energy difference from the unperturbed state, whose energy is the energy of the current (which is infinite due to the infinite domain): 
$$H=K_u[\eta, \Phi_1]+V_u[\eta]+K_l[\eta, \Phi]+V_l[\eta]-\{K_u[0, 0]+V_u[0]+K_l[0, 0]+V_l[0] \},$$
 therefore the Hamiltonian will be evaluated from 
\begin{equation}
\label{mainHam}
\begin{split}
H[\eta, \Phi, \Phi_1]&= \frac{\rho}{2} \int_{\mathbb{R}} \Phi G(\eta, \beta)\Phi dx +\frac{\rho_1}{2} \int_{\mathbb{R}} \Phi_1 G_1(\eta)\Phi_1 dx  \\
&- \rho \int_{\mathbb{R}} (\gamma \eta + \kappa)\Phi(x,t) \eta_x dx + \rho_1 \int_{\mathbb{R}} (\gamma_1 \eta + \kappa)\Phi_1(x,t) \eta_x dx\\
&+ \frac{\rho\gamma^2 - \rho_1 \gamma_1^2 }{6 } \int_{\mathbb{R}}\eta ^3 dx+\frac{g(\rho-\rho_1)+\kappa(\rho \gamma - \rho_1 \gamma_1)}{2} \int_{\mathbb{R}} \eta^2 dx.
\end{split}
\end{equation}
The function above is not yet the Hamiltonian. As the Bernoulli equation \eqref{Bint} suggests, the momentum-type variable, conjugate to the coordinate-type variable $\eta$ is \cite{BB1,BB2,BO}  
\begin{equation} \label{xi}
    \xi:=\rho\Phi-\rho_1 \Phi_1
\end{equation}
Using \eqref{Gdef} and \eqref{G1def} as well as \eqref{kin_interface} we have
\begin{alignat}{2}\label{GG1}
        \left\lbrace
        \begin{array}{lcl}
        G(\eta, \beta)\Phi=-\eta_x({\varphi}_x)_s+ ({\varphi}_y)_s = \eta_t+(\gamma\eta+\kappa)\eta_x,
        \\
        G_1(\eta)\Phi_1=\eta_x({\varphi}_{1,x})_s-({\varphi}_{1,y})_s=-\eta_t-(\gamma_1\eta+\kappa)\eta_x
        \end{array}
        \right.
\end{alignat}
from where 
\begin{alignat}{2} \label{mu}
G(\eta, \beta)\Phi+G_1(\eta)\Phi_1=\mu:=(\gamma-\gamma_1)\eta \eta_x.
\end{alignat}
From \eqref{xi} and \eqref{mu}
\begin{alignat}{2}
\rho_1G(\eta, \beta)\Phi+\rho G_1(\eta)\Phi=\rho_1\mu+ G_1(\eta)\xi.
\end{alignat}
Defining
\begin{alignat}{2}
\label{B_DEF}
B:=\rho G_1(\eta)+\rho_1 G(\eta, \beta)
\end{alignat}
we express $\Phi$ and $\Phi_1$ in terms of $\xi$ and $\eta:$
\begin{alignat}{2}\label{phys}
        \left\lbrace
        \begin{array}{lcl}
        \Phi=B^{-1}\big(\rho_1\mu+G_1(\eta)\xi \big),
        \\
        \Phi_1=B^{-1}\big(\rho\mu-G(\eta, \beta)\xi\big ).
        \end{array}
        \right.
\end{alignat}
Utilizing \eqref{mu}, \eqref{phys} and \eqref{mainHam} we express the Hamiltonian in the form
\begin{multline}
\label{Main_Ham}
H(\eta,\xi)=\frac{1}{2}\int_{\mathbb{R}} \xi G(\eta, \beta) B^{-1}G_1(\eta)\xi \,dx
- \frac{1}{2}\rho\rho_1\int _{\mathbb{R}}  \mu   B^{-1}\mu  \,dx\\
-\int_{\mathbb{R}} (\gamma\eta+\kappa)\xi\eta_x \,dx+\rho_1\int _{\mathbb{R}}\mu B^{-1}G(\eta, \beta)\xi\,dx\\
+ \frac{\rho\gamma^2 - \rho_1 \gamma_1^2 }{6 } \int_{\mathbb{R}}\eta ^3 dx+\frac{g(\rho-\rho_1)+\kappa(\rho \gamma - \rho_1 \gamma_1)}{2} \int_{\mathbb{R}} \eta^2 dx.
\end{multline} 
This expression formally coincides with the expression for the flat bottom \cite{CI2}, apart from the fact that $G(\eta,\beta)$ depends now on the bottom topography. Considering the case where $\gamma_1=\gamma$ and $\mu=0$ and the Hamiltonian reduces further to
\begin{multline}
\label{HamPlisk}
H(\eta,\xi)=\frac{1}{2}\int_{\mathbb{R}} \xi G(\eta, \beta) B^{-1}G_1(\eta)\xi \,dx
-\int_{\mathbb{R}} (\gamma\eta+\kappa)\xi\eta_x \,dx \\
+ \frac{(\rho - \rho_1) \gamma^2 }{6 } \int_{\mathbb{R}}\eta ^3 dx+\frac{(g+\kappa\gamma)(\rho  - \rho_1)}{2} \int_{\mathbb{R}} \eta^2 dx.
\end{multline}
\subsection{Hamiltonian structure}
The Hamiltonian structure for two-layer domains allowing for currents and an internal wave is derived in \cite{Compelli,Compelli2} for the case of a flat bottom. Moreover, the Hamiltonian formulation for the irrotational scenario 
for surface waves over a rough bottom was derived in \cite{CGNS}, developing the perturbative technique for the Dirichlet-Neuman operators for non-even bottom. The currents in the layer $-m\le y \le -l$ do not contribute to the Hamiltonian, as it could be seen from \eqref{K-L2} and \eqref{Main_Ham}. Indeed, as noted in \cite{CompelliIvanov1, Iv17} the motion of the interface is affected only by the motion in the layers adjacent to the interface, so that the equations of motion of the internal wave \eqref{kin_interface} and \eqref{Bint} can be represented in the (non-canonical) Hamiltonian form \begin{equation}
\label{EOMsys}
        \left\lbrace
        \begin{array}{lcl}
        \eta_t=\delta_{\xi} H
        \\
        \xi_t=-\delta_{\eta} H+\Gamma  \chi
        \end{array}
        \right.
\end{equation}
where
\begin{alignat}{2}
\Gamma:=\rho\gamma-\rho_1\gamma_1+2\omega\big(\rho-\rho_1\big)
\end{alignat}
is a constant and
\begin{alignat}{2}
\label{lem2}
\chi(x,t)=- \int_{-\infty}^x\eta_t (x',t)dx'=-\partial_x^{-1}\eta_t
\end{alignat}
is the stream function, evaluated at $y=\eta(x,t),$ (see \cite{Compelli,CIP} for details).

Introducing the variable $\mathfrak{u}=\xi_x$ one can write down \eqref{EOMsys} in the equivalent form
\begin{equation}
\label{EOMsys1}
        \left\lbrace
        \begin{array}{lcl}
        \eta_t=-\left(\delta_{\mathfrak{u}} H\right)_x
        \\
        \mathfrak{u}_t+\Gamma \eta_t=-\left(\delta_{\eta} H\right)_x .
        \end{array}
        \right.
\end{equation}
\begin{rem}
There is a formal transformation of the equations \eqref{EOMsys} into a canonical form
\begin{equation}
        \left\lbrace
        \begin{array}{lcl}
       \eta_t=\delta_{\zeta} H
        \\
        \zeta_t=-\delta_{\eta} H
        \end{array}
        \right.
\end{equation}
by the following change of one of the variables (cf. \cite{Wahlen,Compelli,Compelli2})
\begin{alignat}{2}
\label{vartrans}
\xi\rightarrow\zeta=\xi+\frac{\Gamma}{2} \int_{-\infty}^{x} \eta(x',t)\,dx'.
\end{alignat}
Therefore, the system \eqref{EOMsys}, which could be also represented as  
\begin{equation}
\label{EOMsys2}
        \left\lbrace
        \begin{array}{lcl}
        \eta_t=\delta_{\xi} H
        \\
        \xi_t=-\delta_{\eta} H-\Gamma  \int_{-\infty}^{x} \frac{\delta H}{\delta \xi (x')} dx'
        \end{array}
        \right.
\end{equation}
is Hamiltonian too \cite{Compelli,Compelli2}. 
\end{rem}
The condition
$ \int_{\mathbb{R}} \eta(x',t) dx' = 0$ ensures that 
$$\int _{-\infty}^x \eta(x',t)dx' \in \mathcal{S} (\mathbb{R})$$  and hence $\zeta(x, t) \in \mathcal{S} (\mathbb{R}).$
We note that \begin{align}\frac{d}{dt}H& =\int _{\mathbb{R}}(\delta_{\xi} H \xi_t + \delta_{\eta} H \eta_t) dx =
\int_{\mathbb{R}} \delta_{\xi} H\Gamma \chi dx \nonumber \\
&=-\Gamma \int_{\mathbb{R}} \eta_t \partial_x^{-1} \eta_t dx= -\frac{\Gamma}{2}\int_{\mathbb{R}} \left[\left( \int _{-\infty}^x \eta_t(x',t)dx'\right)^2\right]_x dx =0. 
\end{align}
In what follows we will employ these equations with an approximation for the Hamiltonian functional $H.$

\subsection{Series expansion of the Dirichlet-Neumann operator}\label{DN}
The Hamiltonian \eqref{mainHam} depends on the Dirichlet-Neumann operator
$$G(\eta, \beta)= G_{11}-G_{12}G_{22}^{-1}G_{21}. $$  This is a self-conjugate operator. Some details about $G_{ij}$ at different orders can be found in Appendix \ref{DNO}. The operator can be expanded over the powers of $\eta$ and $\beta$. Let us now introduce appropriate scales. The interfacial waves are assumed of small amplitude, relative to $h$, i.e. $|\eta_{\mathrm{max}}|/h=\varepsilon \ll 1.$  The bottom variations are also considered small, but $|\beta_{\mathrm{max}}|/h $ of order $\tilde{\epsilon} \le \varepsilon^{1/3}.$
The rationale of this choice will become evident later (see the expansion \eqref{DN_1}). We point out that the magnitude of the bottom variations is not fixed by $\e,$ varying within the described limits, with all derivations being valid for the flat bottom as well. In order to keep terms up to the order of $\varepsilon$ we keep in the expansion $\eta^0, \eta^1$ and $\beta^0, \ldots, \beta^3.$  Therefore, in all expansions we are keeping the contributions from the following entries: $$G_{ij} ^{(0,0)}+G_{ij} ^{(1,0)}+G_{ij} ^{(0,1)}+G_{ij} ^{(0,2)}+G_{ij} ^{(0,3)} ,$$ whose orders are explicitly  $$G_{ij} ^{(0,0)}+\varepsilon G_{ij} ^{(1,0)}+\tilde{\epsilon}  G_{ij} ^{(0,1)}+\tilde{\epsilon} ^{2} G_{ij} ^{(0,2)}+\tilde{\epsilon} ^3  G_{ij} ^{(0,3)} .$$  The contribution from $G_{ij} ^{(1,1)} $ is of order $\tilde{\epsilon} \varepsilon $ and is neglected.

The next assumption is the assumption of the slow variations of the bottom profile. Mathematically, we assume that $\beta= \beta(\varepsilon x).$ Then the commutator of $\beta$ and the differentiation operator $D:=-i\partial_x$ is proportional to $\varepsilon \beta'(\varepsilon x)$ which itself is of order $\varepsilon$. Thus, if we keep only terms of order $\varepsilon,$ we can write $\varepsilon^a D \beta \approx \varepsilon^a \beta D $ (where $0<a\le 1$) since the difference $\varepsilon^a D \beta - \varepsilon^a \beta D \sim \varepsilon^{a+1}\ll \varepsilon$ and could be neglected. In other words, with  the exception of the leading order term, we can interchange $\beta D$ and $D \beta$. The expansion involves also the long-wave parameter $\delta=h/\lambda\ll 1$ where $\lambda$ is the typical wavelength. Since $k=2\pi/\lambda$ is the wave number, sometimes we write also symbolically $\delta hk$ instead of $hk$ to remember the fact that the quantity is of order $\delta. $  Moreover, we write $\delta hD$, instead of $hD$ since $k$ is the eigenvalue of $D=-i\partial_x$ when acting on functions representing plane waves $\exp(ikx).$ With these assumptions the truncated expansion is

\begin{equation}
\begin{split}
G(b, \eta)&= \delta^2 D( (h-\beta) + \e \eta )D -\delta^4 D^2\left[\frac{1}{3}(h-\beta)^3+\e h^2 \eta \right] D^2  +  \delta^6 \frac{2}{15}h^5 D^6 \\
& \phantom{*****************************}+\mathcal{O}(\delta^8, \e \delta^6, \e^2 \delta^4)\\
&=\delta^2 D( b(X)+\e \eta)D -\delta^4 D^2 \left[ \frac{1}{3}b^3(X)+\e h^2 \eta \right]D^2 + \delta^6 \frac{2}{15}h^5 D^6+\mathcal{O}(\delta^8, \e \delta^6, \e^2 \delta^4) , \label{DN_1}
\end{split}
\end{equation}
\noindent where $b(X)= h-\beta(\varepsilon x)$ is the local depth and $X=\varepsilon x$ indicates that the bottom depth varies slowly with $x$. This of course coincides with the result from \cite{CIT} which was obtained following a slightly different approach based on the framework from \cite{CGNS}. Assuming that $\mathcal{O}(h)=\mathcal{O}(h_1),$ for the operator $G_1$ we have as usual \cite{CGK}
\begin{multline} 
G_1(\eta)=\delta\Big( D \tanh(\delta h_1 D)\Big)\\- \varepsilon\delta^2\Big(D\eta  D -  D \tanh(\delta h_1 D)  \eta  D \tanh(\delta h_1 D)\Big)+\mathcal{O}(\delta^8, \e \delta^6, \e^2 \delta^4).
\end{multline}
or with the hyperbolic tangent functions expanded, 
\begin{equation}\label{G1}
G_1(\eta)=\delta^2 D\left( h_1 -\varepsilon  \eta \right)D -\delta^4 D^2  \left[ \frac{1}{3}h_1^3-\e h_1^2 \eta \right]D^2   +\mathcal{O}(\delta^8, \e \delta^6, \e^2 \delta^4).
\end{equation}

\section{The long wave approximation}

\noindent We are concerned in this section with the
KdV-like long-wave regime which arises when the relation between the scales is $\varepsilon \sim \delta^2,$ $\xi=\mathcal{O}(\delta),$ and then both $\mathfrak{u}$ and $\eta$ are of order $\delta^2$. Then for the operator $B$ we have 
\begin{equation}
B=\delta^2  D \left( (\rho_1 b(X)+\rho h_1)-\delta^2 \frac{1}{3}D\big(\rho_1  b^3(X)+\rho h_1^3 \big)D
+\varepsilon(\rho_1 -\rho) \eta \right)D+ \mathcal{O}(\delta^6 )
\end{equation} 
which entails that the approximate Hamiltonian whose expansion, including terms of order $\delta^6$, is:
\begin{multline} \label{H}
H(\eta,\mathfrak{u})=\frac{1}{2}\delta^4 \int_{\mathbb{R}} \mathfrak{u}\big ( \alpha_1(\delta^2 x)  + \delta^2 \alpha_3(\delta^2 x)  \eta 
+ \delta^2 \alpha_2(\delta^2 x) \partial_x^2 \big)\mathfrak{u} dx   +\delta^4\alpha_5 \int_{\mathbb{R}} \frac{\eta^2}{2} dx \\
+\delta^4 \kappa \int_{\mathbb{R}} \eta \mathfrak{u} dx +\delta^6 \frac{1}{2}\int \alpha_4(\delta^2 x)  \eta^2 \mathfrak{u} dx 
+\delta^6 \alpha_6\int_{\mathbb{R}}  \frac{\eta^3}{6}dx
\end{multline}
where, using the notation $X=\delta^2 x,$
\begin{equation} \begin{split} \alpha_1(\delta^2 x)&=\frac{b(X) h_1}{\rho_1 b(X)+\rho h_1}, \qquad \alpha_2= \frac{b^2h_1^2(\rho b+\rho_1 h_1)}{3(\rho_1 b+\rho h_1)^2}, \qquad \alpha_3= \frac{\rho h_1^2-\rho_1 b^2}{(\rho_1 b+\rho h_1)^2}, \\
\alpha_4&=\frac{\gamma_1\rho_1 b+ \gamma \rho h_1}{\rho_1 b+\rho h_1}, \quad \alpha_5= g(\rho-\rho_1)+(\rho \gamma -\rho_1 \gamma_1)\kappa, \quad \alpha_6=\rho\gamma^2-\rho_1 \gamma_1^2.
\end{split}\end{equation}
The Hamiltonian equations \eqref{EOMsys1} for the Hamiltonian \eqref{H} in terms of $\eta$ and $\mathfrak{u}=\xi_x$ are (note that the equations written with scales will bring a factor of $\delta^2$ for each Hamiltonian variable, which will compensate the overall factor $\delta^4$ of the Hamiltonian)
\begin{equation}\label{BA0}
\begin{split}
&\eta_t + \kappa \eta_x + \left[\alpha_1 \mathfrak{u} + \delta^2 \alpha_2 \mathfrak{u}_{xx} + \delta^2 \alpha_3 \eta \mathfrak{u}+ \frac{\delta^2}{2}\alpha_4 \eta ^2 \right]_x=0, \\
&\mathfrak{u}_t+\kappa\mathfrak{u}_x +\Gamma \eta_t + \left[ \alpha_5 \eta + \frac{\delta^2}{2} \alpha_3 \mathfrak{u}^2 + \delta^2\alpha_4 \mathfrak{u}\eta + \frac{\delta^2}{2}\alpha_6 \eta^2 \right]_x=0.
\end{split}
\end{equation}
The $x$-derivatives of $\alpha_k(X)$ produce quantities of smaller order. Substituting the explicit form of $\alpha_5$ we have 
\begin{equation}\label{BA01}
\begin{split}
&\eta_t + \kappa \eta_x + (\alpha_1 \mathfrak{u})_x + \delta^2 \alpha_2 \mathfrak{u}_{xxx} + \delta^2 (\alpha_3 (\eta \mathfrak{u})_x+ \alpha_4 \eta \eta_x) =0, \\
&\mathfrak{u}_t+\kappa\mathfrak{u}_x +\Gamma (\eta_t+ \kappa \eta_x)  +(\rho-\rho_1)(g-2\omega \kappa) \eta_x + \delta^2 (\alpha_3 \mathfrak{u}\mathfrak{u}_x + \alpha_4(\mathfrak{u}\eta)_x + \alpha_6 \eta \eta_x)=0.
\end{split}
\end{equation}
Since $\omega=7.3\times 10^{-5}$ rad/s, $\kappa \sim 1 $ m/s, then $g\gg 2 \omega \kappa$ and the $2 \omega \kappa$ term will be neglected.

In the leading order
\begin{equation}\label{lo}
\begin{split}
&\eta_t+\kappa\eta_x+\left(\alpha_1 \mathfrak{u}  \right)_x=0,\\
&\mathfrak{u}_t+\kappa\mathfrak{u}_x +\Gamma(\eta_t+\kappa \eta_x)+(\rho-\rho_1)g \eta_x=0.
\end{split}
\end{equation}
The monochromatic solutions for $\eta$ and $\mathfrak{u}$ can be obtained in the form
\begin{equation}\label{los}
\begin{split}
        \eta(x,t)&=\eta_0e^{ik(x-c(X)t)}\\
\mathfrak{u}(x,t)&=\mathfrak{u}_0e^{ik(x-c(X)t)}
\end{split}
\end{equation}
where $c(X)$ is the wave speed, which depends on the ``slowly varying'' variable $X$. From \eqref{lo} -- \eqref{los} it is straightforward to obtain the following quadratic equation for the wave speed $c$:
\begin{equation}\label{cc}
(c-\kappa)^2 +\alpha_1 \Gamma (c-\kappa)-\alpha_1(\rho-\rho_1)g =0.
\end{equation}
The solutions are
\begin{equation} \label{c}
c(X)=\kappa+\frac{1}{2}\left(-\Gamma \alpha_1(X)\pm\sqrt{\Gamma^2 \alpha_1^2(X) +4g(\rho- \rho_1) \alpha_1(X) }\right).
\end{equation}
For example, for internal waves in the presence of the EUC, taking the typical values $\kappa=1$ m/s, $\gamma_1=-0.1$ s$^{-1}$, $\gamma=0.1$ s$^{-1}$ and depths $h_1=200$ m,  $h=2000$ m  densities $\rho=1037$ kg/m$^3$, $\rho_1=1026,$ kg/m$^3$, 
we have $c_+\approx 5.35$ m/s (right running waves) and $c_-\approx -3.39$ m/s (left running waves). It is evident also that $c(X)$ is of the same order as $\kappa$. Another observation is that the presence of vorticity does not change considerably the wave speed, which for the irrotational case ($\gamma=0,$ $\gamma_1=0$) for example, is  $c_{\pm}\approx 1 \pm 4.37$ m/s. In contrast, the wave speed of the surface waves (whose effect is neglected here) depends significantly on the vorticity near the surface \cite{CIMT}.

As in the previous studies, following \cite{Johnson,Johnson71,CIT}, in addition to the variable $X$, we introduce the characteristic variable in the form
\begin{equation}\label{charvar}
\theta=\frac{1}{\varepsilon}R(X) - t,
\end{equation}
where $R(X)$ is a function such that $R'(X)=1/c(X).$ The $(x,t)$ coordinate partial derivatives change according to
\begin{equation}\label{chvar}
\begin{split}
    \partial_x&\equiv R'(X) \partial_{\theta}+\varepsilon\partial_X=\frac{1}{c(X)}\partial _{\theta}+\varepsilon\partial_X ,\\
    \partial_t&\equiv-\partial_{\theta}.
\end{split}
\end{equation}
The equations then can be transformed from $(x,t)$ variables to the slow variables $(\theta, X).$   This way, of course, two sets of equations arise (for the left and for the right running waves).

The equations written in terms of the new variables are
\begin{equation}\label{eqn1}
\begin{split}
-(c-\kappa)  \eta_{\theta}+ \delta^2 c\kappa\eta_X + \alpha_1 \mathfrak{u}_{\theta}+& \delta^2 c (\alpha_1 \mathfrak{u})_X 
+\delta^2\frac{\alpha_2}{c^2}  \mathfrak{u}_{\theta\theta\theta} \\
&+ \delta^2   \left[ \alpha_3(\mathfrak{u} \eta )_{\theta}+ \alpha_4\eta\eta_{\theta}\right]+\mathcal{O}(\delta^4)  =0 
\end{split}
\end{equation}
and
\begin{multline}
-(c-  \kappa) \mathfrak{u}_{\theta}+ [(\rho- \rho_1)g - \Gamma (c-\kappa)] \eta_{\theta}+\delta^2 c\kappa\mathfrak{u}_X 
+ \delta^2 c[\Gamma \kappa +(\rho-\rho_1)g]\eta_X \\
+\delta^2 [\alpha_3 \mathfrak{u}\mathfrak{u}_{\theta}+ \alpha_4   (\eta\mathfrak{u})_{\theta}
+ \alpha_6 \eta\eta_{\theta} ] +\mathcal{O}(\delta^4)=0.
\end{multline}
From the second equation and \eqref{c}
\begin{multline} \label{u_theta}
\mathfrak{u}_{\theta}
=\frac{c-\kappa}{\alpha_1}  \eta_{\theta}+\delta^2 \frac{\kappa c }{c-\kappa} \mathfrak{u}_X+ \delta^2 \frac{c[\Gamma \kappa + (\rho-\rho_1)g ]}{c-\kappa}\eta_X \\
+\delta^2 \frac{1}{c-\kappa}[\alpha_3 \mathfrak{u}\mathfrak{u}_{\theta}+ \alpha_4   (\eta\mathfrak{u})_{\theta}
+ \alpha_6 \eta\eta_{\theta} ] +\mathcal{O}(\delta^4)=0.
\end{multline}
Therefore, in the leading order we have
\begin{equation}\label{u_eta}
\mathfrak{u}
=\frac{c-\kappa}{\alpha_1}  \eta +\mathcal{O}(\delta^2) .
\end{equation}
Next we substitute \eqref{u_theta} in  \eqref{eqn1} and then we substitute \eqref{u_eta} in the terms of order $\delta^2$ to obtain the single equation for $\eta$

\begin{multline} \label{KdV}
c^2[2(c-\kappa) + \alpha_1 \Gamma ] \eta_X
+ \left[ c^2 c_X -\kappa c (c-\kappa) \frac{\alpha_{1,X}}{\alpha_1} \right]\eta
\\
+\frac{\alpha_2(c-  \kappa)^2 }{\alpha_1 c^2}  \eta_{\theta\theta\theta}  +
 \left[3\frac{\alpha_3}{\alpha_1}(c-\kappa)^2 + 3 \alpha_4 (c-\kappa) + \alpha_1 \alpha_6 \right] \eta\eta_{\theta}
=0,
\end{multline}
which is a KdV-type equation \cite{KdV1895} with variable coefficients that depend on functions, slowly varying with $X.$ From \eqref{cc} one can establish a connection between $c_X$ and $\alpha_{1,X}:$

\begin{equation}
    \alpha_{1,X}=\frac{2(c-\kappa)+\alpha_1 \Gamma}{g(\rho-\rho_1)-\Gamma(c-\kappa)}c_X=\alpha_1 \frac{2(c-\kappa)+\alpha_1 \Gamma}{(c-\kappa)^2}c_X. 
\end{equation}


One possible limit of equation \eqref{KdV} is the irrotational case where $\omega=\gamma=\gamma_1=0.$ Hence, it follows that $\Gamma=\alpha_4=\alpha_6=0$,  $c^2=g(\rho-\rho_1)\alpha_1$ and thus, equation \eqref{KdV} acquires the form 
\begin{equation}\label{KdV0}
    \eta_X + \frac{b_X}{4b(X)}\frac{\rho h_1}{\rho_1 b + \rho h_1}\eta + \frac{\rho b + \rho_1 h_1}{6cg(\rho-\rho_1)} \eta_{\theta\theta\theta} + \frac{3c(\rho h_1^2 - \rho_1 b^2)}{2g(\rho-\rho_1)b^2 h_1^2}\eta \eta_{\theta}=0,\end{equation}
which showed up in non-dimensional form in \cite{DR2} lacking, however, details of its derivation.

The limit of \eqref{KdV} to one layer of fluid which coincides to the lower domain $\Omega$ (Fig. \ref{fig:thesisfigure_systemG}) corresponds to $\rho_1=0. $  Then $\Gamma=\rho(\gamma+ 2 \omega),$ $\alpha_1=b(X)/\rho,$ $\alpha_2=b^3(X)/(3\rho),$ $\alpha_3=1/\rho,$ $\alpha_4=\gamma,$ $\alpha_6=\rho \gamma^2$ and
\begin{equation} 
c(X)=\kappa+\frac{1}{2}\left(-(\gamma+2\omega) b(X)\pm\sqrt{(\gamma+2\omega)^2 b^2(X) +4gb(X) }\right).
\end{equation}
Then the equation \eqref{KdV} becomes
\begin{multline} \label{KdV1}
c^2[2(c-\kappa) + b (\gamma+ 2 \kappa) ] \eta_X
+ \left[ c^2 c_X -\kappa c (c-\kappa) \frac{b_{X}}{b } \right]\eta
\\
+\frac{b^2(c-  \kappa)^2 }{3  c^2}  \eta_{\theta\theta\theta}  +
 \left[\frac{3}{b}(c-\kappa)^2 + 3 \gamma  (c-\kappa) + b \gamma^2  \right] \eta\eta_{\theta}
=0.
\end{multline}
This situation corresponds to surface waves over one layer of fluid and has been analysed in \cite{CIMT}. Further reduction of \eqref{KdV1} could be obtained for the irrotational case with $\kappa=0,$ $b=c^2/g.$ The equation acquires the form
\begin{equation} \label{KdV000}
 (2c  \eta_X+ c_X \eta)+\frac{c^2}{3g^2  }  \eta_{\theta\theta\theta}  + \frac{3g }{c^2}  \eta\eta_{\theta}=0,
\end{equation}
which is the equation derived by Johnson \cite{Johnson71,Johnson} for one layer of irrotational fluid over variable bottom; we refer the reader to \cite{CIT} for its derivation by means of a Hamiltonian approach. Therefore equations \eqref{KdV}, \eqref{KdV1} provide generalisations of Johnson's equation.
These equations resemble the KdV equation, the important distinction being that they exhibit variable coefficients. While the integrability of KdV type systems is well established in numerous investigations for a long time \cite{ZMNP}, it seems that these particular models are, for a general choice of $b(X),$ not integrable \cite{OV} .

\section{Fission of solitons moving over a step-like bottom}

\begin{figure}
\centering
\begin{subfigure}[t]{0.46\textwidth}
\centering
\includegraphics[width=\textwidth]{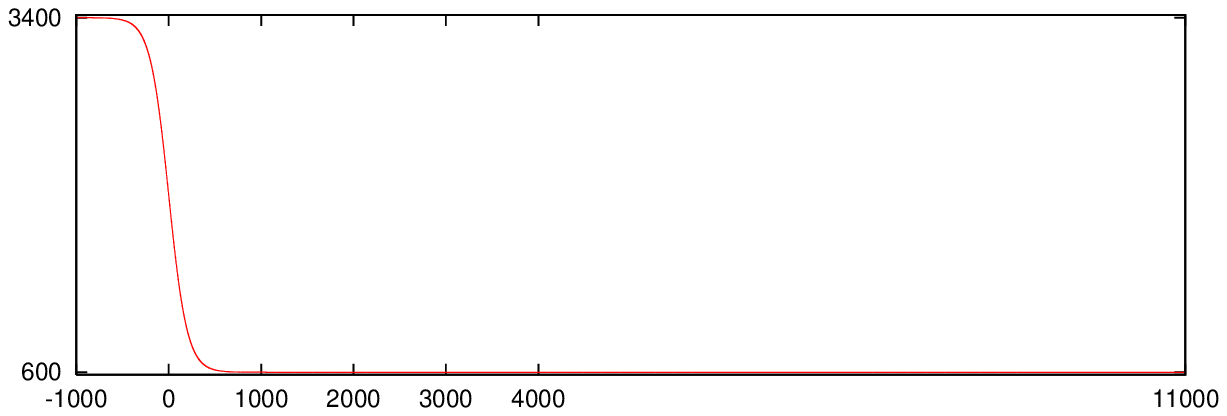}
\caption{Local depth profile, $\alpha=0.7$, $\tilde{\beta}=10$ - from deep (3400 m) to shallow (600 m). The horizontal axis is for the variable $X.$} \label{fig:2a}
\end{subfigure}
   \hfill
\begin{subfigure}[t]{0.46\textwidth}
\centering
\includegraphics[width=\textwidth]{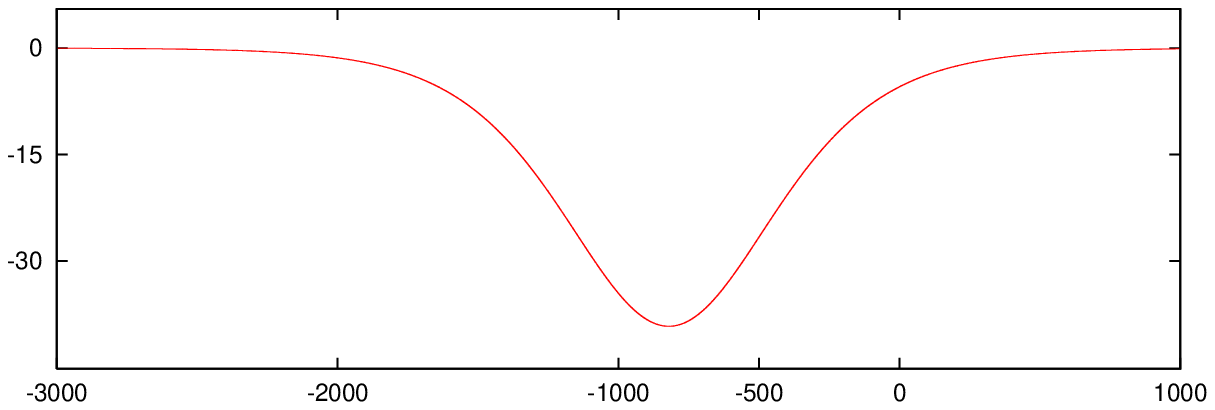}
\caption{In this case  the initial condition is a one-soliton solution with $K= 0.002,$ $\omega=\gamma=\gamma_1=0.$ The horizontal axis is for the variable $\theta.$}\label{fig:2b}
\end{subfigure}
\caption{Sketch of the local depth and the soliton profile of the initial condition,  $K=0.002$. }
\end{figure}

We study an example where the step-like bottom is modeled by $$b(X)=h\left[1-\alpha \tanh\left(\tilde{\beta}\frac{X}{h}\right )\right ].$$ The constants are taken as the actual values in the SI system of units as follows: $h=2000$, $h_1=200,$ $g=9.81$ $\rho=1037$, $\rho_1=1026.$ The constants $\alpha,\tilde{\beta}$ are given in each case. First, we study the irrotational case \eqref{KdV0}. The bottom threshold is located at $X=0$, the initial condition is an exact KdV soliton coming from $X\to -\infty$ where the depth is $b\to b_0=h(1+\alpha).$ When $\alpha>0$ the soliton moves from deep to shallow regions, when $\alpha<0$ - from shallow to deep. The depth profile is given on Fig. \ref{fig:2a}.

\noindent When $\kappa=0,$ considering right-moving waves, we have 

\begin{equation} \label{c_r}
c(X)=\frac{1}{2}\left(-\Gamma \alpha_1(X)+\sqrt{\Gamma^2 \alpha_1^2(X) +4g(\rho- \rho_1) \alpha_1(X) }\right),
\end{equation}
and
\begin{equation} \label{cX}
c_X=-\frac{\alpha \tilde{\beta }\rho  \alpha_1}{[2c(X)+\alpha_1 \Gamma]b^2\cosh ^2 (\tilde{\beta}\frac{X}{h})} [c(X)]^2.
\end{equation}
The initial condition is the soliton that satisfies the unperturbed KdV equation with $b=b_0=h(1+\alpha):$
\begin{multline} \label{KdV00}
c_0^2[2c_0 + \alpha_{1,0} \Gamma ] \eta_X
+\frac{\alpha_{2,0}}{\alpha_{1,0} }  \eta_{\theta\theta\theta}  +
 \left[3\frac{\alpha_{3,0}}{\alpha_{1,0}}c_0^2 + 3 \alpha_{4,0} c_0  + \alpha_{1,0} \alpha_6 \right] \eta\eta_{\theta}
=0
\end{multline}
where all quantities with sub-index $0$ are evaluated with $b=b_0=h(1+\alpha),$  including $c_0=[c(X)]_{b=b_0}:$


\begin{equation} \label{c0}
c_0=\frac{1}{2}\left(-\Gamma \alpha_{1}+\sqrt{\Gamma^2 \alpha_{1}^2 +4g(\rho- \rho_1) \alpha_{1} }\right)_{b=b_0}.
\end{equation}
Writing the previous KdV equation \eqref{KdV00} in the form
\begin{equation} \label{KdV-2}
        \eta_X + \mathcal{A}_0\eta \eta_\theta +\mathcal{B}_0 \eta_{\theta \theta \theta} =0,
\end{equation}
where  $\mathcal{A}_0= (\mathcal{A})_{b=b_0},$  $\mathcal{B}_0= (\mathcal{B})_{b=b_0},$

$$ \mathcal{A}:=\frac{3\frac{\alpha_{3}}{\alpha_{1}}[c(X)]^2 + 3 \alpha_{4} c(X)  + \alpha_{1} \alpha_6}{[c(X)]^2[2c(X) + \alpha_{1} \Gamma ]}, \qquad \mathcal{B} := \frac{\frac{\alpha_{2}}{\alpha_{1} } }{[c(X)]^2[2c(X) + \alpha_{1} \Gamma ]} $$
are constants, we note that it is well known that it has an one-soliton solution
    \begin{equation} \label{KdV-lim}\eta(X,\theta)= \frac{12\mathcal{B}_0}{\mathcal{A}_0}\cdot \frac{K ^2  }{\cosh^2 [K (\theta-\theta_0- 4 K^2 \mathcal{B}_0  X)]},  \end{equation}
where $K, \theta_0$ are arbitrary constants. Therefore, we can take an initial condition at $X=X_0<0$ (ideally modelling a soliton coming from $X\to-\infty$ ) which is an exact solution of the equation for $X\to-\infty$:
 \begin{equation} \label{KdV-lim-final}\eta(X_0,\theta)= \frac{12\mathcal{B}_0}{\mathcal{A}_0}\cdot \frac{K ^2  }{\cosh^2 [K (\theta-\theta_0- 4 K^2 \mathcal{B}_0  X_0)]} . \end{equation} Actually for the choice of $X_0$ it is sufficient that $\tanh(\tilde{\beta} X_0/h) \approx -1$, e.g. $|\tilde{\beta }X_0 /h |> 2.5$ ) noting that $X$ plays the role of a ``time'' variable.  
The initial soliton profile is given on Fig. \ref{fig:2b}.
 
 The explanation of the soliton fission when the initial soliton \eqref{KdV-lim-final} reaches the threshold at $X=0$ follows from the Quantum Mechanical theory of the P\"oschl-Teller potential $\text{sech}^2(K\theta),$ cf. \cite{Fl}. The Lax operator for the KdV-equation has the form of a Schr\"odinger equation type spectral problem for the eigenfunction $\psi(\theta)$ with potential $ \eta(\theta,0)=\mathcal{E}K^2\text{sech}^2(K\theta),$  see the details for example in \cite{ZMNP}. Then, since the spectral problem is iso-spectral, that is $X$-independent, we note the following. The eigenfunction $\psi$ corresponds to a potential (KdV-solution) with $N$ discrete eigenvalues if $\mathcal{E}=-N(N+1)/2.$  This corresponds also to an $N$-soliton KdV solution. Let us now take an initial condition at $X_0$ such that $\eta(\theta,X_0)$ is the one-soliton ($N=1$) solution for $X_0\to -\infty,$  and $N>1$ soliton solution for $X\to \infty.$ Then taking into account the constant coefficients of the corresponding KdV equations in the two situations, (when $X\to \pm \infty$) we have an equality of the coefficient in front of the sech$^2$- potential, which is $X$-independent,
 \begin{equation} \label{NN}
 \frac{1(1+1)}{2}  \left(\frac{12\mathcal{B}}{\mathcal{A}}\right)_{b=h(1+\alpha)} K^2= \frac{N(N+1)}{2}  \left(\frac{12\mathcal{B}}{\mathcal{A}}\right)_{b=h(1-\alpha)}K^2
 \end{equation} therefore
  \begin{equation}\label{ratio}
    \frac{N(N+1)}{2}= \left(\frac{\mathcal{B}}{\mathcal{A}}\right)_{b=h(1+\alpha)} / \left(\frac{\mathcal{B}}{\mathcal{A}}\right)_{b=h(1-\alpha)} .
 \end{equation}
Some more details are available for example in \cite{CIT}. 

Introducing the notation $b^*=h(1-\alpha)$ in the irrotational case \eqref{KdV0} we have
   \begin{equation} \label{NN1}
    \frac{N(N+1)}{2}=\frac{b_0}{b^*} \cdot \frac{\rho_1h_1+\rho b_0}{\rho_1h_1+\rho b^*} \cdot  \frac{\rho_1b_0+\rho h_1}{\rho_1 b^*+\rho h_1}\cdot \frac{\rho h_1^2-\rho_1 (b^*)^2}{\rho h_1^2-\rho_1 b_0^2}.
 \end{equation}
  This formula gives $N=2.07$ while in reality we observe at least 3 solitons.
  
\begin{figure}
 \centerline{\includegraphics[width= .80\textwidth]{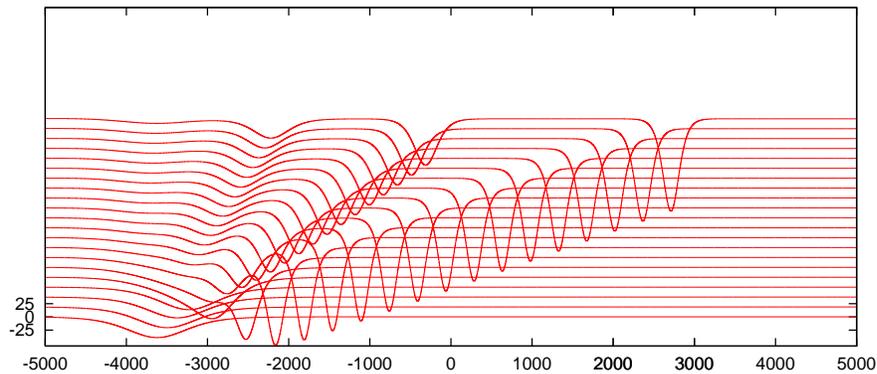}}
\caption{Soliton fission under the conditions on Fig. \ref{fig:2a} and \ref{fig:2b}, $-1000<X<6000$,  $\omega=\gamma=\gamma_1=0.$ The horizontal axis is for the variable $\theta.$ The waterfall plot corresponds to increasing values of the variable $X.$}
\label{fig:3}
\end{figure}

The numerical solution is presented at Fig. \ref{fig:3}. Actually, formula \eqref{NN1} could be improved by noticing that with an integrating factor, introducing $F(\theta,X)=\sqrt{c(X)}\eta(\theta,X)$ the equation \eqref{KdV0} transforms into an equation for $F$ in an exact KdV form.
Since $\sqrt{c(X)}\sim [\alpha_1(X)]^{1/4}$ the formula \eqref{NN1} acquires an extra factor 
\[ \left([\alpha_1(X)]^{1/4} \right)_{b=b_0}/ \left([\alpha_1(X)]^{1/4} \right)_{b=b^*}
\]
 then we obtain the formula 
 \begin{equation} \label{NN2}
    \frac{N(N+1)}{2}=\left(\frac{b_0}{b^*}\right)^{5/4}  \frac{\rho_1h_1+\rho b_0}{\rho_1h_1+\rho b^*}   \left(\frac{\rho_1b_0+\rho h_1}{\rho_1 b^*+\rho h_1}\right)^{3/4}  \frac{\rho h_1^2-\rho_1 (b^*)^2}{\rho h_1^2-\rho_1 b_0^2}.
 \end{equation}
 which also appears in \cite{DR2} where it is obtained from the arguments of Johnson \cite{Johnson71}, see also \cite{CIT}. In particular, for our data it gives $N=2.14.$  The discrepancy could be due to the fact that the derivation of \eqref{NN} is based on the equality of the KdV-amplitudes at the moment of the hitting of a step-like threshold. The threshold in our case however is not sharp, it is modelled by the smooth tanh function. In addition, over the region of the obstacle the equations are not exactly KdV equations, since their coefficients depend in general on the bottom variations in the region of the smooth threshold.  Moreover, at the obstacle there is always a reflected wave, which is not taken into account. So the formula for $N$ should be considered only as an estimate.
 
\begin{figure}
 \centerline{\includegraphics[width= .80\textwidth]{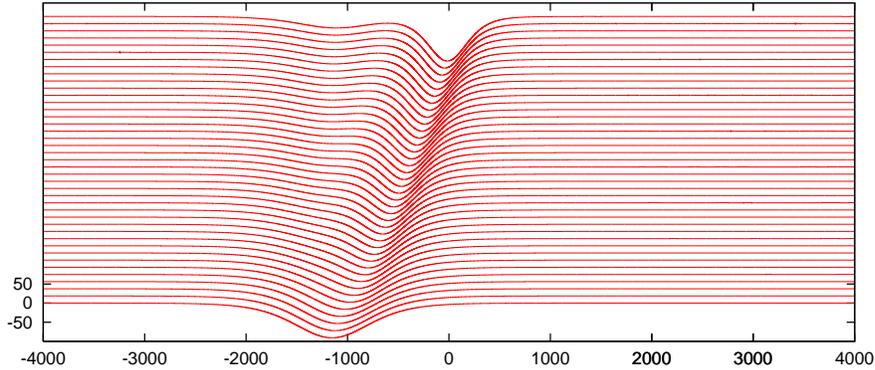}}
\caption{ Soliton fission with nonzero vorticity: $\alpha=0.7$, $\tilde{\beta}=10$, $\gamma=0.1,$ $\gamma_1=-0.1$, $K=0.002$, $-1000<X<11000$.}
\label{fig:4}
\end{figure}

\begin{figure}
\centering
\begin{subfigure}[t]{0.42\textwidth}
\centering
\includegraphics[width=\textwidth]{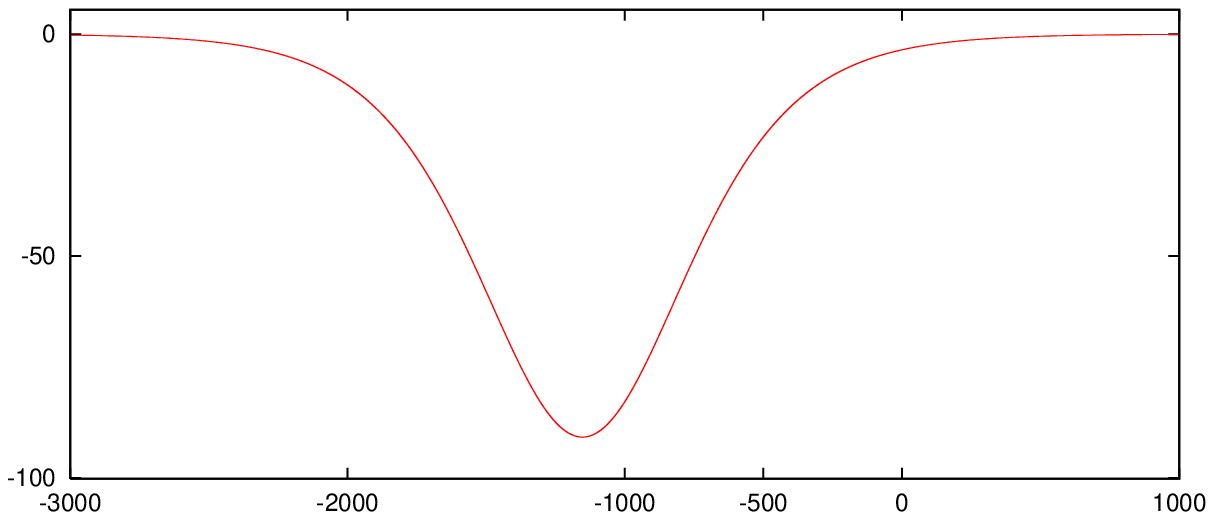}
\caption{The initial shape of the soliton from Fig.\ref{fig:4}. } \label{fig:5a}
\end{subfigure}
   \hfill
\begin{subfigure}[t]{0.56\textwidth}
\centering
\includegraphics[width=\textwidth]{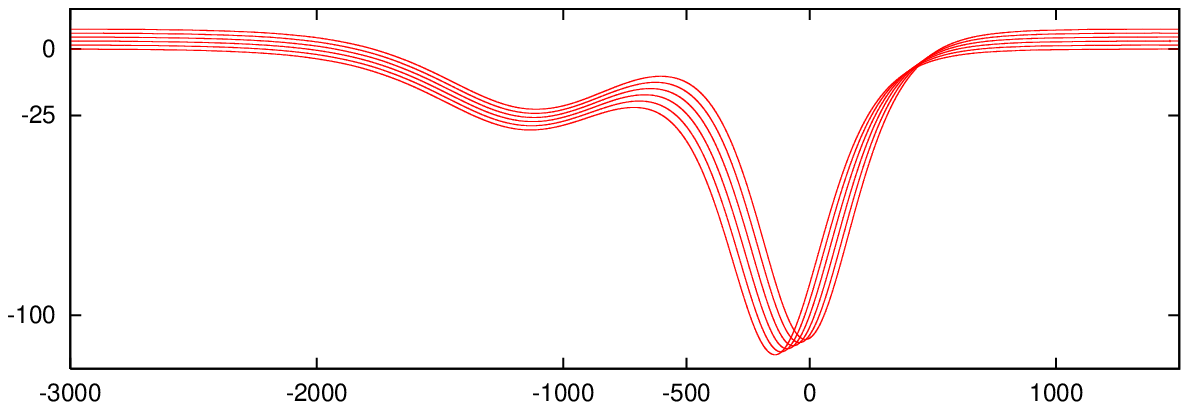}
\caption{The last 6 slices of the soliton evolution from Fig.\ref{fig:4}.  }\label{fig:5b}
\end{subfigure}
\caption{ Soliton fission with nonzero vorticity: the initial and final stages of the process from Fig.\ref{fig:4}.}
\end{figure}

Next, we study numerically the situation with nonzero vorticities. We take $\gamma=0.1,$ $\gamma_1=-0.1$ and $\omega=0,$ being very small in comparison to the other vorticities. The depth $b(X)$-profile is the same as on Fig. \ref{fig:2a} however when $K=0.002,$ the amplitude of depression of the initial condition rises to 100 meters (which is not unusual for internal waves) as shown on Fig. \ref{fig:5a}. The soliton fission is less pronounced giving two solitons, as it could be seen from Fig. \ref{fig:4} and Fig. \ref{fig:5b}.

\begin{figure}
\centering
\begin{subfigure}[t]{0.45\textwidth}
\centering
\includegraphics[width=\textwidth]{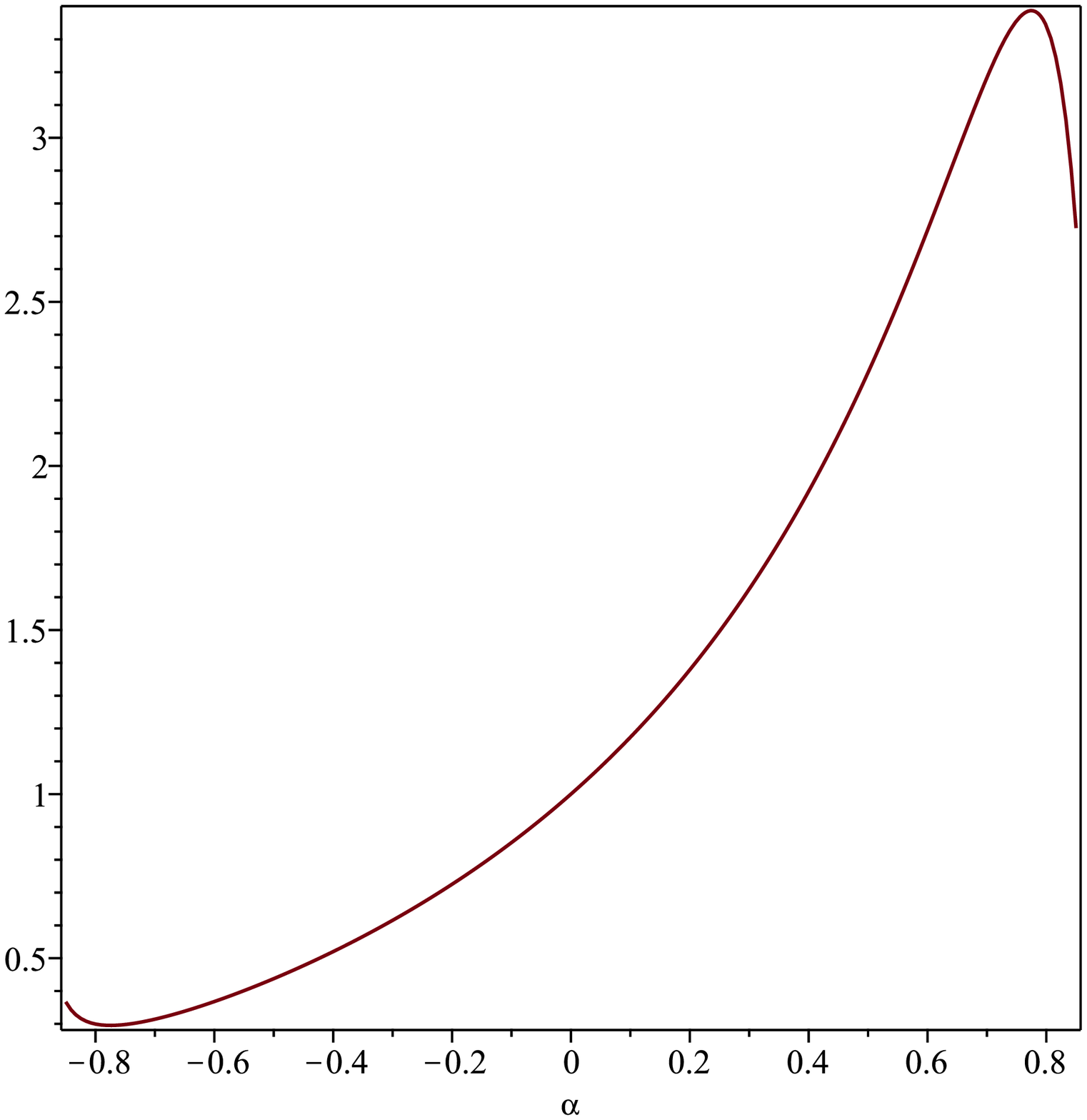}
\caption{The ratio \eqref{ratio} in the irrotational case.  } \label{fig:6a}
\end{subfigure}
   \hfill
\begin{subfigure}[t]{0.45\textwidth}
\centering
\includegraphics[width=\textwidth]{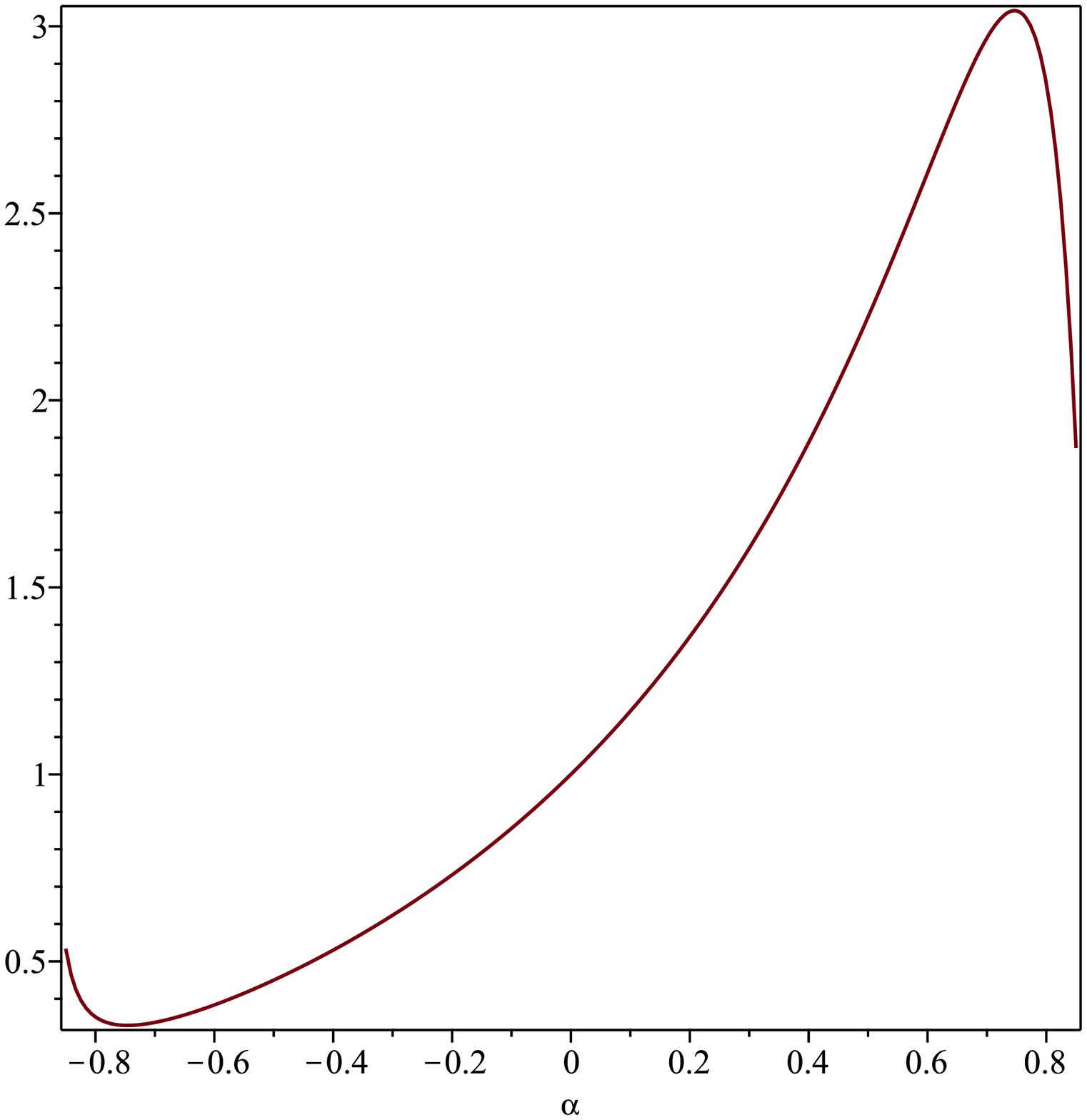}
\caption{The ratio \eqref{ratio} when vorticities are  $\gamma=0.1,$ $\gamma_1=-0.1.$  }\label{fig:6b}
\end{subfigure}
\caption{ The dependence of the ratio \eqref{ratio} on $\alpha,$ giving $N(N+1)/2$ as a function of the threshold magnitude $\alpha.$ }
\end{figure}

The approximate ratio \eqref{ratio} gives $N\approx 1.98,$ which is in an agreement with the results. One could hope for an improved formula like in the irrotational case, however the integration factor is apparently not in a simple form and we are going to limit ourselves with the approximation \eqref{ratio}. We provide the dependence of the ratio \eqref{ratio} on the step magnitude $\alpha$  on Fig. \ref{fig:6a} and \ref{fig:6b}. It is evident that in both cases the maximum of the ratio is reached for $\alpha \approx 0,7$  which is already quite an extreme value. In the rotational case the ratio $N(N+2)/2$ barely reaches the value of $3$ which corresponds to just two solitons, which are actually observed in the numerical experiments. The findings show that the solitons of the internal waves are quite robust (in comparison to those of the surface waves, \cite{CIT,CIMT}) and remain stable for relatively mild bottom variations. 

In the case $\alpha<0,$ as predicted from the theory, no soliton fission is observed, the initial soliton very slowly decays loosing its energy through waves of radiation in the region $X>0.$

\section{Special case when the coefficient of the nonlinear term is of smaller order or vanishing}

There is a particular case when the coefficient 
\begin{equation} \label{coef_2}
     3\frac{\alpha_3}{\alpha_1}(c-\kappa)^2 + 3 \alpha_4 (c-\kappa) + \alpha_1 \alpha_6 
\end{equation}
of the $\eta \eta_{\theta}$ term in \eqref{KdV},
is of order of  $\e$ or smaller. In fact, there are situations with values of $X$ and the parameters such that this coefficient could be zero or very closed to zero, so that the main nonlinearity term is $\eta^2 \eta_{\theta}.$  This could be seen immediately in the irrotational case  \eqref{KdV0} when $\rho_1 b^2= \rho h_1^2. $

In this section we explore the situation with small or vanishing coefficient \eqref{coef_2}. 
Since the dispersive term $\eta_{\theta \theta \theta}$ matches the order of the nonlinear term $\eta^2 \eta_{\theta},$  this indicates that the scaling is $\eta$ and $\mathfrak{u}$ of order $\delta,$  that is $\varepsilon $  and $\delta$ of the same order. Since we have taken $X=\delta^2 x,$ where the scale of the wave variations is over $\delta x$, we have now $X=\e^2 x.$ The variable changes to the characteristic variables like in  \eqref{charvar}, \eqref{chvar} become
\begin{equation}\label{charvar1}
\begin{split}
\theta&=\frac{1}{\varepsilon ^2}R(X) - t,\\
\partial_x&\equiv \frac{1}{c(X)} \partial_{\theta}+\varepsilon ^2 \partial_X\notag\\
 \partial_t&\equiv-\partial_{\theta}.
\end{split}
\end{equation}
The extended Hamiltonian with terms of leading order $\e^2$ up to order $\e ^4$ could be obtained from \eqref{Main_Ham}  
and the expansions \eqref{DN_1} and \eqref{G1} with $\delta=\e$
\begin{multline} \label{H11}
H(\eta,\mathfrak{u})=\frac{1}{2}\e^2 \int_{\mathbb{R}} \mathfrak{u}\big ( \alpha_1(\e ^2x)  + \e \alpha_3(\e^2 x)  \eta 
+ \e^2 \alpha_2(\e^2 x) \partial_x^2 \big)\mathfrak{u} dx   +\e ^2\alpha_5 \int_{\mathbb{R}} \frac{\eta^2}{2} dx \\
+\e ^2 \kappa \int_{\mathbb{R}} \eta \mathfrak{u} dx +\e ^3 \frac{1}{2}\int \alpha_4(\e^2 x)  \eta^2 \mathfrak{u} dx 
+\e ^3 \alpha_6\int_{\mathbb{R}}  \frac{\eta^3}{6}dx \\
-\e ^4 \frac{1}{2} \int_{\mathbb{R}} \beta_1(\e^2 x) \eta^2 \mathfrak{u}^2 dx -\e^4 \int_{\mathbb{R}} \beta_2(\e^2 x) \eta^3 \mathfrak{u} dx
- \e^4 \frac{1}{2}\int_{\mathbb{R}} \beta_3(\e^2 x) \frac{\eta^4}{4}  dx,
\end{multline}
where the new terms of order $\e ^4$ are with coefficients
\begin{equation} \begin{split} 
\beta_1(\e^2 x)&=\frac{\rho \rho_1(b(X)+ h_1)^2}{(\rho_1 b(X)+\rho h_1)^3}, \qquad \beta_2= \frac{\rho \rho_1 (\gamma - \gamma_1)( b+ h_1)}{3(\rho_1 b+\rho h_1)^2}, \qquad \beta_3= \frac{\rho \rho_1 (\gamma- \gamma_1)^2}{\rho_1 b+\rho h_1}, 
\end{split}
\end{equation}

The two Hamiltonian equations lead to the analogues of \eqref{eqn1}  and \eqref{u_theta}, however with extra terms, arising from the contributions of the terms with $\beta_1, \beta_2$ and $\beta_3 $ in the Hamiltonian:  
\begin{equation}\label{eqn1a}
\begin{split}
-(c-\kappa)  \eta_{\theta}+  \alpha_1 \mathfrak{u}_{\theta} + \e   \left( \alpha_3 \mathfrak{u} \eta + \alpha_4 \frac{\eta^2}{2} \right)_{\theta} & + \e ^2 c\kappa\eta_X + \e ^2  c (\alpha_1 \mathfrak{u})_X 
+\e ^2 \frac{\alpha_2}{c^2}  \mathfrak{u}_{\theta\theta\theta} \\
&- \e^2(\beta_1 \eta^2  \mathfrak{u} + \beta_2 \eta^3)_{\theta} +\mathcal{O}(\e ^3)  =0, 
\end{split}
\end{equation}
\begin{multline} \label{u_theta_a}
\mathfrak{u}_{\theta}
=\frac{c-\kappa}{\alpha_1}  \eta_{\theta}  +\e  \frac{1}{c-\kappa} \left(\frac{1}{2}\alpha_3 \mathfrak{u}^2 + \alpha_4   \eta\mathfrak{u} + \frac{1}{2}\alpha_6 \eta^2 \right)_{\theta}+\e^2 \frac{\kappa c }{c-\kappa} \mathfrak{u}_X+ 
\e ^2 \frac{c \alpha_5}{c-\kappa}\eta_X \\
-\e^2 \frac{1}{c-\kappa}\left( \beta_1 \eta \mathfrak{u}^2 + 3 \beta_2 \eta^2 \mathfrak{u}+\frac{1}{2}\beta_3 \eta^3 \right)_{\theta} +\mathcal{O}(\e ^3)=0.
\end{multline}
From \eqref{u_theta_a} we have the following relation in the leading order
\begin{equation} \label{u_lo}
    \mathfrak{u}=\frac{c-\kappa}{\alpha_1}  \eta +   \mathcal{O}(\e),
\end{equation}
and from \eqref{u_theta_a} and \eqref{u_lo} we obtain

\begin{equation} \label{u_fo}
    \mathfrak{u}=\frac{c-\kappa}{\alpha_1}  \eta +\e  \left(\frac{\alpha_3(c-\kappa)}{2\alpha_1^2}+ 
    \frac{\alpha_4}{\alpha_1}    + \frac{\alpha_6}{2(c-\kappa)} \right)\eta ^2+   \mathcal{O}(\e^2).
\end{equation}
We substitute \eqref{u_theta_a} in \eqref{eqn1a} and then in the so obtained equation we eliminate $\mathfrak{u}$ with the help of \eqref{u_fo}, thus obtaining the following equation for $\eta:$

\begin{multline} 
\e^2 c^2[2(c-\kappa) + \alpha_1 \Gamma ] \eta_X
+ \e^2 \left[ c^2 c_X -\kappa c (c-\kappa) \frac{\alpha_{1,X}}{\alpha_1} \right]\eta
+\e^2 \frac{\alpha_2(c-  \kappa)^2 }{\alpha_1 c^2}  \eta_{\theta\theta\theta} \\ +
 \e \left[3\frac{\alpha_3}{\alpha_1}(c-\kappa)^2 + 3 \alpha_4 (c-\kappa) + \alpha_1 \alpha_6 \right] \eta\eta_{\theta}\\
+\e^2\left[(2\alpha_3(c-\kappa)+\alpha_1 \alpha_4)\left(\frac{\alpha_3(c-\kappa)}{2\alpha_1^2}+ 
    \frac{\alpha_4}{\alpha_1}    + \frac{\alpha_6}{2(c-\kappa)} \right) \right.\\ \left. -\frac{2\beta_1(c-\kappa)^2}{\alpha_1}-4\beta_2(c-\kappa)-\frac{\alpha_1 \beta _3}{2}\right](\eta^3)_{\theta}=0,
\end{multline}
Finally, if the order of the coefficient \eqref{coef_2} itself is of order $\e$ or smaller, we observe that all terms are of the same order, hence giving a mKdV-type equation  

\begin{multline} \label{mKdV}
 c^2[2(c-\kappa) + \alpha_1 \Gamma ] \eta_X
+ \left[ c^2 c_X -\kappa c (c-\kappa) \frac{\alpha_{1,X}}{\alpha_1} \right]\eta
+ \frac{\alpha_2(c-  \kappa)^2 }{\alpha_1 c^2}  \eta_{\theta\theta\theta} \\ +
  \left[3\frac{\alpha_3}{\alpha_1}(c-\kappa)^2 + 3 \alpha_4 (c-\kappa) + \alpha_1 \alpha_6 \right] \eta\eta_{\theta}\\
+\left[(2\alpha_3(c-\kappa)+\alpha_1 \alpha_4)\left(\frac{\alpha_3(c-\kappa)}{2\alpha_1^2}+ 
    \frac{\alpha_4}{\alpha_1}    + \frac{\alpha_6}{2(c-\kappa)} \right) \right.\\ \left. -\frac{2\beta_1(c-\kappa)^2}{\alpha_1}-4\beta_2(c-\kappa)-\frac{\alpha_1 \beta _3}{2}\right](\eta^3)_{\theta}=0,
\end{multline}

The equation \eqref{mKdV} generalises \eqref{KdV}. Indeed, it works in the scaling used in the derivation of \eqref{KdV} as well, because in this scaling the term with $\eta^3$ could be neglected. Other authors also suggest the mKdV-type equation as a suitable generalisation avoiding the problem with the vanishing term in front of the $\eta^2$-coefficient, \cite{Mad,Grim10}. Finally we point out that the mKdV equation with constant coefficients is integrable, \cite{ZMNP} so that, one can embark on developing soliton perturbation theory for \eqref{mKdV}. 

\section{Conserved quantities}

With an integrating factor 
\begin{equation}
I(X)=\frac{1}{ c^2[2(c-\kappa) + \alpha_1 \Gamma ]}\exp\left(\int _{-\infty}^X \frac{\left[ c^2 c_X -\kappa c (c-\kappa) \frac{\alpha_{1,X}}{\alpha_1} \right]}{ c^2[2(c-\kappa) + \alpha_1 \Gamma ]} dX'\right)
\end{equation}
the equation \eqref{mKdV} acquires the following form 
\begin{equation} \label{mKdVmc}
    E_X + [\tilde{P}(X) E_{\theta \theta}+ \tilde{Q}(X) E^2 + \tilde{R}(X) E^3]_{\theta}=0,
\end{equation}
for the quantity 
\begin{equation}
\begin{split}
E(X,\theta)&:=I(X) c^2[2(c-\kappa) + \alpha_1 \Gamma ] \eta(X,\theta)\\
&=\eta(X, \theta) \exp\left(\int _{-\infty}^X \frac{\left[ c^2 c_X -\kappa c (c-\kappa) \frac{\alpha_{1,X}}{\alpha_1} \right]}{ c^2[2(c-\kappa) + \alpha_1 \Gamma ]} dX'\right).
\end{split}
\end{equation}
We note that for $\Gamma=\kappa =0,$  the above relationship is just $E=\sqrt{c(X)}\eta.$
The conserved quantity (mass conservation) from \eqref{mKdVmc} is 
\begin{equation}
    \int _{\mathbb{R}}E(X,\theta) d\theta :=m_0=\text{const},
\end{equation}
Hence \begin{equation}
    \int _{\mathbb{R}}\eta(X,\theta) d\theta =m_0 \exp\left[-\left(\int _{-\infty}^X \frac{\left[ c^2 c_X -\kappa c (c-\kappa) \frac{\alpha_{1,X}}{\alpha_1} \right]}{ c^2[2(c-\kappa) + \alpha_1 \Gamma ]} dX'\right)\right].
\end{equation}
The interpretation of this type of results is not straightforward - see the comments in the variable bottom section of the Johnson's book \cite{Johnson}. The mass conservation makes sense for the full system and not just for the solution describing the pycnocline.  Multiplying \eqref{mKdVmc} by $E$ leads further to 
\begin{equation}\label{mKdVec}
    \left(\frac{E^2}{2}\right)_X + \left[\tilde{P}(X) E E_{\theta \theta}-\tilde{P}\frac{E^2_{\theta}}{2}+ \frac{2\tilde{Q}(X)}{3} E^3 + \frac{3\tilde{R}(X)}{4} E^3 \right]_{\theta}=0,
\end{equation}
and therefore we obtain an analogue of the ''energy'' conservation, which reads as
\begin{equation}
    \frac{1}{2}\int _{\mathbb{R}}E^2(X,\theta) d\theta =\text{const}.
\end{equation}

\section{Conclusions}
The main achievement of this paper is the derivation of a KdV type equation \eqref{KdV} which describes the propagation of interfacial internal waves in two-layer domains bounded below by a variable bottom and above by a flat surface.
Our analysis includes the shear currents in the two domains and hinges on a consistent derivation of the Dirichlet-Neumann (DN) operators in the Boussinesq approximation. While the final result for the bottom-dependent DN operator \eqref{DN_1}  recovers the one from Craig et al. \cite{CGNS}, the setup of its derivation opens up new possibilities towards an application to a multi-layer system of fluids, which will be explored in forthcoming publications. The bottom-dependent DN operator allows the application of the ''nearly''-Hamiltonian formulation, developed for the configuration of two fluid layers in \cite{Compelli2,CI2} following the DN approach of Craig et al. \cite{CGK}. 

An example for a possible realistic situation are the equatorial waves and currents in the equatorial Pacific Ocean, where the so-called Equatorial Undercurrent resides and where \emph{the abyssal hills} are the most abundant seabed structures near the equator. These Pacific Ocean hills are typically 50-300 m in height, with a width of 2-5 km and a length of 10–20 km \cite{Dil}. Other seabed structures are the seamounts which are higher, but with horizontal dimensions of the same order, that is, bottom structures with horizontal diameters of 2-20 km are typical. Since the bottom length scale over wavelength ratio is of order $\sim 1/\delta, $ which could be a factor of 2-10, the modelling setup is a realistic scenario for waves of 0.5-5 km wavelength.

The general equation \eqref{KdV} in various limits leads to several known simplified cases with variable bottom, like the irrotational case \eqref{KdV0}, the single layer case with background current \eqref{KdV1} and without current \eqref{KdV000}, which goes back to the well-known work of Johnson \cite{Johnson71}.  
In addition, it has been noted that for some values of the parameters the coefficient of the nonlinear term of the model equation might be close to zero, rendering the next order term $\eta^2 \eta_{\theta}$ of increased significance. The scaling for this special case is identified and a model equation of mKdV type is derived \eqref{mKdV}, utilising the Hamiltonian method.  

Wave-breaking of solitary waves is another very significant and interesting topic \cite{VH}. However, its analytical studies will require modelling beyond the KdV and Boussinesq-type models.

\section{Acknowledgements}
\noindent R.I. is partially supported by the Bulgarian National Science Fund, grant K$\Pi$ -06H42/2 from 27.11.2020. C.I.M. acknowledges the support of the Austrian Science Fund (FWF) through research grant P 33107-N. The authors are thankful to two anonymous referees for their numerous suggestions which have improved the text of the article. 

\appendix
\section{}

\subsection{Dirichlet-Neumann operators for variable bottom}  \label{DNO}

In this section we derive the set of Dirichlet-Neumann operators for the lower layer, which is bounded below by a variable bottom. 
Recalling that the bottom is given as $y=-h+\beta(x)=\mathcal{B}(x)$
we define
\begin{equation}
\begin{pmatrix}
G_{11} & G_{12}\\
G_{21} & G_{22}
\end{pmatrix} \begin{pmatrix}\Phi\\ \Phi_b\end{pmatrix}=
\begin{pmatrix}
(\nabla\varphi)_s\cdot {\bf n}_s\sqrt{1+\eta_x^2}\\
(\nabla\varphi)_b\cdot {\bf n}_b\sqrt{1+\beta_x^2}
\end{pmatrix}
\end{equation}
where ${\bf n}_s=(-\eta_x,1)/\sqrt{1+\eta_x^2}$ and ${\bf n}_b=(\beta_x,-1)/\sqrt{1+\beta_x^2}$ are the outward unit normal vectors corresponding to the interface and to the bottom, respectively.

\begin{figure}[ht!]
\begin{center}
\fbox{\includegraphics[totalheight=0.23\textheight]{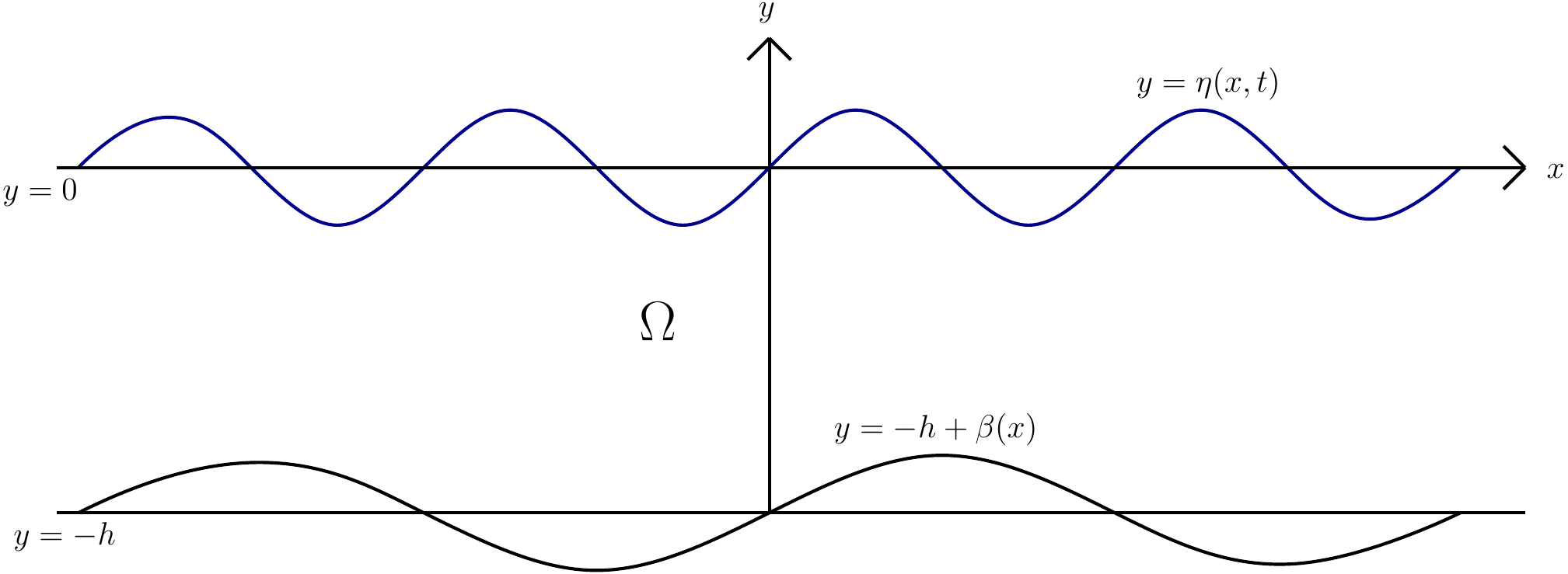}}
\caption{The domain of the lower layer.}
\label{fig:thesisfigure_systemG}
\end{center}
\end{figure}
\noindent Therefore, the last equality is written as
\begin{equation}
\begin{pmatrix}
G_{11} & G_{12}\\
G_{21} & G_{22}
\end{pmatrix} \begin{pmatrix}\Phi\\ \Phi_b\end{pmatrix}=
\begin{pmatrix}
(\nabla\varphi)_s\cdot (-\eta_x, 1)\\
-(\nabla\varphi)_b\cdot (-\beta_x, 1)
\end{pmatrix}.
\end{equation}

\noindent In what follows we work with a fixed wave number $k$ and a special associated harmonic function $\varphi_k(x,y)=(a(k)e^{ky}+b(k)e^{-ky})e^{ikx}.$ The corresponding values on the surface and at the bottom are\footnote{The coefficient $b(k)$ is not related to the function $b(X)$ from the previous sections.}
$$\Phi_{k}(x)=(ae^{k\eta(x)}+b e^{-k\eta(x)})e^{ikx},\quad \Phi_{b,k}(x)=(a e^{-k(h-\beta(x))}+b e^{k(h-\beta(x))})e^{ikx}$$
Expanding we have (summation over $j\ge 0$)
\begin{equation}
\Phi_{k}=\sum\f{1}{j!}\eta^j k^j(a+(-1)^jb)e^{ikx}
\end{equation}
and
\begin{equation}
\Phi_{b,k}=\sum\f{1}{j!}\beta^j k^j( a e^{-kh}+(-1)^j b e^{kh})e^{ikx}
\end{equation}
Since $\varphi_k(x,y)=(ae^{ky}+be^{-ky})e^{ikx}$ it follows that
\begin{equation}
\begin{split}
(\nabla\varphi_k)_s\cdot (-\eta_x, 1)=&\Big[-ik\eta_x(ae^{k\eta}+be^{-k\eta})+k(ae^{k\eta}
-be^{-k\eta})\Big]e^{ikx}\\
=\Big[-ik\eta_x &\sum\f{1}{j!}(a+(-1)^j b)(k\eta)^j+k\sum\f{1}{j!}(a-(-1)^jb)(k\eta)^j\Big] e^{ikx}
\end{split}
\end{equation}
Likewise
\begin{equation}
\begin{split}
(\nabla\varphi_k)_b\cdot (-\beta_x, 1)
=\Big[ik\beta_x &\sum\f{1}{j!}(a e^{-kh}+ b (-1)^j e^{kh})(k\beta)^j\\
&-k\sum\f{1}{j!}(a e^{-kh}-(-1)^j b e^{kh})(k\beta)^j\Big] e^{ikx}
\end{split}
\end{equation}
\begin{equation}
\begin{split}
&\begin{pmatrix}
G_{11} & G_{12}\\
G_{21} & G_{22}
\end{pmatrix}
\begin{pmatrix}
\sum\f{1}{j!}\eta^j k^j(a+(-1)^jb)e^{ikx}\\
\sum\f{1}{j!}\beta^j k^j(a e^{-kh}+(-1)^j b e^{kh})e^{ikx}
\end{pmatrix}\\
&=\begin{pmatrix}
\Big[-ik\eta_x \sum\f{1}{j!}(a+(-1)^j b)(k\eta)^j+k\sum\f{1}{j!}(a-(-1)^jb)(k\eta)^j\Big] e^{ikx}\\
\Big[ik\beta_x \sum\f{1}{j!}((-1)^j a e^{-kh}+ b e^{kh})(k\beta)^j-k\sum\f{1}{j!}( a e^{-kh}-(-1)^j b e^{kh})(k\beta)^j\Big] e^{ikx}
\end{pmatrix}
\end{split} \label{GenExp}
\end{equation}
We assume that each entry $G_{mn}$  could be written as an infinite series $$G_{mn}=\sum_{p\ge 0,q\ge0}G_{mn}^{(p,q)}(\eta,\beta)$$ where $G_{mn}^{(p,q)}(\eta,\beta)$
is a homogeneous expression of its arguments in a sense that for any two constants $C_1,C_2$

$$G_{mn}^{(p,q)}(C_1 \eta,C_2\beta)= C_1^p C_2^q G_{mn}^{(p,q)}(\eta,\beta).$$ The coefficients $a(k)$ and $b(k)$ in the expressions above are arbitrary, therefore we can equate the corresponding coefficients in \eqref{GenExp} and we have
\begin{equation}
\begin{split}
G_{11}^{(0,0)}+G_{12}^{(0,0)}e^{-kh}=k\\
G_{11}^{(0,0)}+G_{12}^{(0,0)}e^{kh}=-k,
\end{split}
\end{equation}
which implies that
$$G_{11}^{(0,0)}=D\coth(h D),\,\,G_{12}^{(0,0)}=-D\csch(h D).$$
Similarly, we get that
\begin{equation}
\begin{split}
G_{21}^{(0,0)}+G_{22}^{(0,0)}e^{-kh}=-ke^{-kh}\\
G_{21}^{(0,0)}+G_{22}^{(0,0)}e^{kh}=ke^{kh},
\end{split}
\end{equation}
that delivers the solution operators
$$G_{21}^{(0,0)}=-D\csch(h D),\,\,G_{22}^{(0,0)}=D\coth(h D).$$
Thus, the homogeneous component of order zero of $G$ is
\begin{equation}
G^{(0,0)}=
\begin{pmatrix}
D\coth(h D) & -D\csch(h D)\\
-D\csch(h D) & D\coth(h D)
\end{pmatrix}
\end{equation}
We proceed to compute now the operator $G^{(1,0)}$, that is the one which exhibits powers of type $\eta^1$ and
$\beta^0$. First, we have
\begin{equation}
\begin{split}
&G_{11}^{(0,0)}(\eta k)(a-b)e^{ikx}+G_{11}^{(1,0)}(a+b)e^{ikx}+G_{12}^{(1,0)}(ae^{-kh}+be^{kh})e^{ikx}\\
&=[-ik\eta_x(a+b)+k(a+b)\eta k]e^{ikx}.
\end{split}
\end{equation}
Hence
\begin{equation}
\begin{split}
G_{11}^{(1,0)}+G_{12}^{(1,0)}e^{-kh}=-ik\eta_x +\eta k^2-G_{11}^{(0,0)}\eta k\\
G_{11}^{(1,0)}+G_{12}^{(1,0)}e^{kh}=-ik\eta_x +\eta k^2+G_{11}^{(0,0)}\eta k
\end{split}
\end{equation}
from where we get
$$G_{11}^{(1,0)}e^{ikx}=\big[-ik\eta_x +\eta k^2-G_{11}^{(0,0)}(\eta k)\coth(kh)\big]e^{ikx},$$
which means that
$$G_{11}^{(1,0)}=D\eta D-D\coth(hD)\eta D\coth(hD).$$
Likewise
$$G_{12}^{(1,0)}e^{ikx}=G_{11}^{(0,0)}(\eta k)\frac{1}{\sinh(kh)}e^{ikx},$$
that is
$$G_{12}^{(1,0)}=D\coth(hD)(\eta D)\csch(hD).$$
Furthermore
\begin{equation}
\begin{split}
G_{21}^{(1,0)}+G_{22}^{(1,0)}e^{-kh}=-G_{21}^{(0,0)}(\eta k)\\
G_{21}^{(1,0)}+G_{22}^{(1,0)}e^{kh}=G_{21}^{(0,0)}(\eta k)
\end{split}
\end{equation}
which implies that
\begin{equation}
\begin{split}
G_{22}^{(1,0)}=G_{21}^{(0,0)}(\eta k)\frac{1}{\sinh(kh)},\\
G_{21}^{(1,0)}=-G_{22}^{(1,0)}\cosh(kh).
\end{split}
\end{equation}
Thus, in operatorial form we have
\begin{equation}
\begin{split}
G_{22}^{(1,0)}=-D\csch(hD)(\eta D)\csch(hD),\\
G_{21}^{(1,0)}=D\csch(hD)(\eta D)\coth(hD).
\end{split}
\end{equation}
Summarizing, we have
\begin{equation}
G^{(1,0)}=\begin{pmatrix}
D\eta D-D\coth(hD)\eta D\coth(hD) & D\coth(hD)(\eta D)\csch(hD)\\
D\csch(hD)(\eta D)\coth(hD) & -D\csch(hD)(\eta D)\csch(hD)
\end{pmatrix}
\end{equation}
We proceed with $G^{(0,1)}$. First we have
\begin{equation}
 \begin{split}
  G_{11}^{(0,1)}(a+b)+G_{12}^{(0,1)}(ae^{-kh}+be^{kh})+G_{12}^{(0,0)}(\beta k)(ae^{-kh}-be^{kh})=0,
 \end{split}
\end{equation}
that implies
\begin{equation}
 \begin{split}
  G_{11}^{(0,1)}+G_{12}^{(0,1)}e^{-kh}=&- G_{12}^{(0,0)}(\beta k)e^{-kh}\\
  G_{11}^{(0,1)}+G_{12}^{(0,1)}e^{kh}=& G_{12}^{(0,0)}(\beta k)e^{kh}.
 \end{split}
\end{equation}
Hence,
\begin{equation}
 \begin{split}
G_{12}^{(0,1)}=-D\csch(hD)(\beta D)\coth(hD)\\
G_{11}^{(0,1)}=D\csch(hD) (\beta D)\csch(hD)
\end{split}
\end{equation}
Now we use
\begin{equation}
 \begin{split}
  G_{21}^{(0,1)}(a+b)+G_{22}^{(0,0)}(\beta k)(-ae^{-kh}+be^{kh})+G_{22}^{(0,1)}(ae^{-kh}+be^{kh})\\
  =ik\beta_x (ae^{-kh}+be^{kh})-k(ae^{-kh}+be^{kh})\beta k,
 \end{split}
\end{equation}
that delivers
\begin{equation}
 \begin{split}
G_{21}^{(0,1)}+G_{22}^{(0,1)}e^{-kh}=\left(-G_{22}^{(0,0)}\beta k+ik\beta_x-\beta k^2\right) e^{-kh}\\
G_{21}^{(0,1)}+G_{22}^{(0,1)}e^{kh}=\left(G_{22}^{(0,0)}\beta k+ik\beta_x-\beta k^2\right)e^{kh}.
 \end{split}
\end{equation}
It follows from above that
\begin{equation}
 \begin{split}
 G_{22}^{(0,1)}=& G_{22}^{(0,0)}(\beta k)\coth(kh)-D\beta D\\
 =& D\coth(hD)(\beta D)\coth(hD)-D\beta D
 \end{split}
\end{equation}

Likewise we have $G_{21}^{(0,1)}(e^{kh}-e^{-kh})=-2G_{22}^{(0,0)}(\beta k)$, that is
\begin{equation}
 \begin{split}
  G_{21}^{(0,1)}=-G_{22}^{(0,0)}(\beta D)\frac{1}{\sinh(hD)}=-D\coth(hD)(\beta D)\csch(hD).
 \end{split}
\end{equation}
Summarizing, we have
\begin{equation}
 G^{(0,1)}=\begin{pmatrix}
  D\csch(hD) (\beta D)\csch(hD) &- D\csch(hD)(\beta D)\coth(hD)\\
  -D\coth(hD)(\beta D)\csch(hD) & \coth(hD)(\beta D)\coth(hD)-D\beta D
 \end{pmatrix}
\end{equation}
To compute $G^{(0,2)}$ we notice
\\
\begin{equation}
\begin{split}
&G_{11}^{(0,2)}(a+b)+G_{12}^{(0,2)}(ae^{-kh}+be^{kh})+G_{12}^{(0,1)}(\beta k)(ae^{-kh}-be^{kh})\\
&+G_{12}^{(0,0)}(\frac{1}{2}\beta^2 k^2)(ae^{-kh}+be^{kh})=0
\end{split}
\end{equation}
\\
from where it follows that
\\
\begin{equation}
\begin{split}
G_{11}^{(0,2)}+G_{12}^{(0,2)}e^{-kh}=-G_{12}^{(0,1)}(\beta k)e^{-kh}-G_{12}^{(0,0)}(\frac{1}{2}\beta^2 k^2)e^{-kh}\\
G_{11}^{(0,2)}+G_{12}^{(0,2)}e^{kh}=G_{12}^{(0,1)}(\beta k)e^{kh}-G_{12}^{(0,0)}(\frac{1}{2}\beta^2 k^2)e^{kh}.
\end{split}
\end{equation}
\\
Solving above we find that
\\
\begin{equation}
\begin{aligned}
G_{11}^{(0,2)}\sinh(kh)&=-G_{12}^{(0,1)}(\beta k)\\
G_{12}^{(0,2)}&=G_{12}^{(0,1)}(\beta k)\coth(kh)-G_{12}^{(0,0)}(\frac{1}{2}\beta^2 k^2),
\end{aligned}
\end{equation}
\\
so that, in operatorial form we have
\\
\begin{equation}
\begin{aligned}
G_{11}^{(0,2)}&=\csch(hD)(D\beta )(D\coth(hD))(\beta D)\csch(hD)\\
G_{12}^{(0,2)}&=-D\csch(hD)\left[(\beta D)\coth(hD)\right]^2+D\csch(hD)\left(\frac{1}{2}\beta^2 D^2\right).
\end{aligned}
\end{equation}
\\
  Furthermore,
\\
\begin{equation}
\begin{aligned}
&G_{21}^{(0,2)}(a+b)+G_{22}^{(0,2)}(ae^{-kh}+be^{kh})+G_{22}^{(0,1)}(\beta k)(a e^{-kh}-be^{kh})\\
&+G_{22}^{(0,0)}(\frac{1}{2}\beta^2 k^2)(ae^{-kh}+be^{kh})=ik^2\beta\beta_x (ae^{-kh}-be^{kh}) -k^3\beta^2(ae^{-kh}-be^{kh}),
\end{aligned}
\end{equation}
\\
that is,
\\
\begin{equation}
\begin{aligned}
G_{21}^{(0,2)}+G_{22}^{(0,2)}e^{-kh}=\left(-G_{22}^{(0,1)}(\beta k)-G_{22}^{0,0}(\frac{1}{2}\beta^2 k^2)
+ik^2\beta\beta_x- \frac{k^3\beta^2}{2} \right)e^{-kh}\\
G_{21}^{(0,2)}+G_{22}^{(0,2)}e^{kh}=\left(G_{22}^{0,1}(\beta k)-G_{22}^{0,0}(\frac{1}{2}\beta^2 k^2)
-ik^2\beta\beta_x +\frac{k^3\beta^2}{2}\right)e^{kh}.
\end{aligned}
\end{equation}
\\
From the above system we derive
\\
\begin{equation}
\begin{aligned}
G_{21}^{(0,2)}=\left(G_{22}^{(0,1)}(\beta k)+i \beta\beta_x k^2-\frac{\beta^2 k^3}{2}\right)\csch(kh)\\
G_{22}^{(0,2)}=\left(-G_{22}^{(0,1)}(\beta k)-i \beta\beta_x k^2 +\frac{\beta^2 k^3}{2}\right)\coth(kh)-G_{22}^{(0,0)}(\frac{1}{2}\beta^2 k^2).
\end{aligned}
\end{equation}
\\
We would like to note now that the operator corresponding to $i\beta\beta_x k^2-\frac{\beta^2 k^3}{2}$ is
$$i\partial_x \left(\frac{\beta^2}{2}\right)\cdot D^2-\frac{\beta^2 D^3}{2}=-D\left( \frac{\beta^2}{2}D^2\right)
+\frac{\beta^2}{2} D^3-\frac{\beta^2}{2}D^3=-\frac{1}{2}D\left(\beta^2 D^2\right)$$
\\
Hence, we have
\\
\begin{equation}
\begin{aligned}
G_{21}^{(0,2)}=&\left(-D(\coth(hD)(\beta D))^2 +D(\beta D)^2 -\frac{1}{2}D\left(\beta^2 D^2\right)\right)\csch(hD)\\
G_{22}^{(0,2)}=&\left(D(\coth(hD)(\beta D))^2-D(\beta D)^2 +\frac{1}{2}D\left(\beta^2 D^2\right)\right)\coth(hD)\\
&-D\coth(hD)(\frac{1}{2}\beta^2 D^2).
\end{aligned}
\end{equation}
\\
Availing of the identity
$$D(\beta D)^2=\frac{1}{2}(D (\beta^2 D^2)+D^2(\beta^2 D))$$
we obtain that
\begin{equation}
\begin{aligned}
G_{21}^{(0,2)}=&\left(-D(\coth(hD)(\beta D))^2 +\frac{1}{2}D^2(\beta^2 D)\right)\csch(hD)\\
=&-[\coth(hD) (D\beta)]^2(D\csch (hD)+\frac{1}{2}D^2\beta^2 D\csch(hD)
\end{aligned}
\end{equation}
and
\begin{equation}
\begin{aligned}
G_{22}^{(0,2)}=& D[\coth(hD)(\beta D)]^2 \coth(hD)\\
&-\frac{1}{2}D^2(\beta^2 D)\coth(hD)   -D\coth(hD)(\frac{1}{2}\beta^2 D^2)\\
=& \coth(hD)(D\beta) [D\coth(hD)]\beta D)\coth(hD)\\
&-\frac{1}{2}D^2(\beta^2 D)\coth(hD)   -D\coth(hD)(\frac{1}{2}\beta^2 D^2)
\end{aligned}
\end{equation}

It is now evident that $G_{12}^{(0,2)}$ and $G_{21}^{(0,2)}$ are conjugate to each other and $G_{11}^{(0,2)}$ and $G_{21}^{(0,2)}$ are self-conjugate operators.

To compute the $G^{(0,3)}$ terms we write first
\begin{equation}
\begin{aligned}
&G_{11}^{(0,3)}(a+b)+G_{12}^{(0,1)}(\frac{1}{2}\beta^2 k^2)(ae^{-kh}+be^{kh})+G_{12}^{(0,2)}(\beta k)(ae^{-kh}-be^{kh})\\
&+G_{12}^{(0,3)}(ae^{-kh}+be^{kh})+G_{12}^{(0,0)}\frac{(\beta k)^3}{6}(ae^{-kh}-be^{kh})=0,
\end{aligned}
\end{equation}
from which we conclude
\begin{equation}
\begin{aligned}
&G_{11}^{(0,3)}+G_{12}^{(0,3)}e^{-kh}=\left(-G_{12}^{(0,2)}(\beta k)-G_{12}^{(0,1)}(\frac{1}{2}\beta^2 k^2)  - G_{12}^{(0,0)}\frac{(\beta k)^3}{6}  \right) e^{-kh}\\
&G_{11}^{(0,3)}+G_{12}^{(0,3)}e^{kh}=\left(  G_{12}^{(0,2)}(\beta k)-G_{12}^{(0,1)}(\frac{1}{2}\beta^2 k^2)  + G_{12}^{(0,0)}\frac{(\beta k)^3}{6}  \right) e^{kh},
\end{aligned}
\end{equation}
that implies
$$ G_{12}^{(0,3)}=G_{12}^{(0,0)}\frac{(\beta D)^3}{6}\coth(hD)-G_{12}^{(0,1)}(\frac{1}{2}\beta^2 D^2)+G_{12}^{(0,2)}(\beta D)\coth(hD),  $$
\begin{equation}
\begin{aligned}
G_{12}^{(0,3)}&=-D\csch(hD)[\beta D\coth(hD)]^3 \\
&+\csch(hD) (D\beta )\left[ \left(\frac{1}{2}\beta D^2 \beta  -\frac{1}{6}\beta^2 D^2 \right)D\coth(hD)+\frac{1}{2}D \coth(hD) \beta^2 D^2 \right].\\
\end{aligned}
\end{equation}

and
$$ G_{11}^{(0,3)}=-\left(G_{12}^{(0,2)}(\beta D)+\frac{1}{6}G_{12}^{(0,0)}\beta^3 D^3\right)\csch(h D),   $$
\begin{equation}
\begin{aligned}
G_{11}^{(0,3)}=&D\csch(hD)(\beta D)[\coth(hD) \beta \coth(hD)] (D\beta )(D\csch (hD)) \\
&-D\csch(hD)\left(\frac{1}{2}\beta^2 D^2\beta -\frac{1}{6}\beta^3 D^2 \right)D\csch(hD).
\end{aligned}
\end{equation}
\\

Although it is not obvious, the operator $\frac{1}{2}\beta^2 D^2\beta -\frac{1}{6}\beta^3 D^2$ is self-conjugate, see the identity \eqref{Id1}, thus $G_{11}^{(0,3)}$ is a self-conjugate operator.

To find the entries in the second row of the matrix $G^{(0,3)}$ we write
\\
\begin{equation}
\begin{aligned}
G_{21}^{(0,3)}(a+b)+&G_{22}^{(0,1)}(\frac{1}{2}\beta^2 k^2)(ae^{-kh}+be^{kh})+G_{22}^{(0,2)}(\beta k)(ae^{-kh}-be^{kh})\\
+&G_{22}^{(0,3)}(ae^{-kh}+be^{kh})+G_{22}^{(0,0)}(\frac{\beta^3 k^3}{6})(ae^{-kh}-be^{kh})\\
&=ik\beta_x(\beta k)^2\frac{ae^{-kh}+be^{kh}}{2}
-k(\beta k)^3\frac{ae^{-kh}+be^{kh}}{6}.
\end{aligned}
\end{equation}
\\
Hence,
\\
\begin{equation}
\begin{aligned}
G_{21}^{(0,3)}+ & G_{22}^{(0,3)}e^{-kh}\\
&= \left( -G_{22}^{(0,1)}(\frac{1}{2}\beta^2 k^2)-G_{22}^{(0,2)}(\beta k)
-G_{22}^{(0,0)}(\frac{\beta^3 k^3}{6}) +\frac{ik\beta_x (\beta k)^2}{2}-\frac{k(\beta k)^3}{6}\right)e^{-kh}\\
G_{21}^{(0,3)}+ & G_{22}^{(0,3)}e^{kh}\\
&=\left( -G_{22}^{(0,1)}(\frac{1}{2}\beta^2 k^2)+G_{22}^{(0,2)}(\beta k)
+G_{22}^{(0,0)}(\frac{\beta^3 k^3}{6})
+\frac{ik\beta_x (\beta k)^2}{2}-\frac{k(\beta k)^3}{6}\right) e^{kh},
\end{aligned}
\end{equation}
\\
\begin{equation}
\begin{aligned}
G_{21}^{(0,3)}=-\left(G_{22}^{(0,2)}(\beta D) +G_{22}^{(0,0)}(\frac{\beta^3 D^3}{6})\right) \csch(hD)
\end{aligned}
\end{equation}
\\
\begin{equation}
\begin{aligned}
G_{21}^{(0,3)}&=-[\coth(hD) (D \beta)]^3D\csch(hD) \\
&+\left[ D \coth(h D) \left(\frac{1}{2}\beta^2 D^2   -\frac{1}{6}\beta^3 D^2 \frac{1}{\beta}\right)+\frac{1}{2}D^2 \beta^2 D \coth(hD)  \right]\beta D \csch(hD) ,\\
\end{aligned}
\end{equation} with the identity \eqref{Id1} we obtain
\begin{equation}
\begin{aligned}
G_{21}^{(0,3)}&=-[\coth(hD) (D \beta)]^3D\csch(hD) \\
&+\left[ D \coth(h D) \left(\frac{1}{2}\beta D^2 \beta   -\frac{1}{6} D^2 \beta^2 \right)+\frac{1}{2}D^2 \beta^2 D \coth(hD)  \right]\beta D \csch(hD) ,\\
\end{aligned}
\end{equation}

We observe that $G_{21}^{(0,3)}$ is conjugate to $G_{12}^{(0,3)}$.
\\
Finally, noting that $$ \left(\frac{ik\beta_x (\beta k)^2}{2}-\frac{k(\beta k)^3}{6}\right)e^{ikx}= -\frac{1}{6} D \beta^3 D^3 e^{ikx}$$ we obtain
\begin{equation}
\begin{aligned}
G_{22}^{(0,3)}= G_{22}^{(0,0)}\frac{1}{6}\beta^3 D^3 \coth(hD) -G_{22}^{(0,1)}(\frac{1}{2}\beta^2 D^2)+G_{22}^{(0,2)}(\beta D)\coth(hD)-\frac{1}{6}D \beta^3 D^3
\end{aligned}
\end{equation}

\begin{equation}
\begin{aligned}
G_{22}^{(0,3)}=& \coth(hD) (D \beta) \coth(hD) (D\beta D ) \coth(hD) (\beta D) \coth(hD) \\
&-\frac{1}{2}D^2 \beta^2 D \coth(hD) \beta D \coth (hD) -\frac{1}{2} D \coth (hD) \beta D \coth(hD) \beta^2 D^2 \\
&-\frac{1}{2} D \coth(hD) \left[\beta^2 D^2 \beta - \frac{1}{3} \beta^3 D^2  \right] D \coth(hD)\\
 &+\frac{1}{2} D  \left[\beta D \beta^2 D - \frac{1}{3} \beta^3 D^2  \right] D
\end{aligned}
\end{equation}

The operators in the square brackets above are self-conjugate. This could be checked using identities like \eqref{Id1}, \eqref{Id2}. Hence, both operators $G_{11}^{(0,3)}$ and $G_{22}^{(0,3)}$ are self-conjugate, $(G_{12}^{(0,3)})^{*}=G_{21}^{(0,3)}$ and the whole matrix - valued operator $ G^{(0,3)}$ is self-conjugate.

\subsection{Identities}

The conjugation of an operator $\mathcal{A}$ is with respect to the inner product  $$ (f,g)=\int_{\mathbb{R}} \bar{f}(x) g(x) dx. $$  The definition of $\mathcal{A}^*$ is $$ (\mathcal{A}^*f, g)= (f, \mathcal{A} g) $$ for any choice of $f,g \in \mathcal{S}(\mathbb{R}).$
We note that the physically meaningful variables are real and the operators, involved in the Hamiltonian description are invariant under complex conjugation.

We have the following identities which show that the corresponding operators are self-conjugate:

\begin{equation}\label{Id1}
 \beta^2 D^2 \beta -\frac{1}{3} \beta^3 D^2 = \beta D^2 \beta^2 -\frac{1}{3}D^2 \beta^3
\end{equation}

\begin{equation}\label{Id2}
 \beta D \beta^2 D -\frac{1}{3} \beta^3 D^2 = D\beta^2 D \beta -\frac{1}{3}D^2 \beta^3
\end{equation}

The proof of these identities relies on the fact that the commutator between $D$ and a function like $\beta(x)$ is  $ D \beta - \beta D = -i\beta_x.$

\section{} \label{numap}

\subsection{Finite-difference implementation of KdV-type equation with variable coefficients}
In this appendix we describe briefly the numerical scheme for solving the main KdV-type equation with variable coefficients \eqref{KdV} for the function $\eta=\eta(X,\theta):$
\begin{align}
\tilde{A}(X)\eta_X + \tilde{B}(X)\eta+\tilde{C}(X)\eta_{\theta\theta\theta}+\tilde{D}(X)\eta\eta_{\theta}=0\end{align}
with  $X$-dependent variable coefficients
\begin{equation}
    \begin{split}
        \tilde{A}(X)&=c^2 [2(c-\kappa)+\alpha_1 \Gamma],\\
        \tilde{B}(X)&=\left[c^2 c_X-\kappa c (c-\kappa)\frac{\alpha_{1,X}}{\alpha_1} \right],\\
        \tilde{C}(X)&=\frac{\alpha_2(c-\kappa)^2}{\alpha_1 c^2},\\
        \tilde{D}(X)&=\left[3\frac{\alpha_3}{\alpha_1}(c-\kappa)^2+3\alpha_4(c-\kappa)+\alpha_1 \alpha_6 \right].
    \end{split}
\end{equation}
We assume an initial condition of the form of one-soliton solution \eqref{KdV-lim-final}.

We consider an uniform mesh in the interval $[-L_1, L_2]$ $\theta_i = (i - 1)\Delta\theta$, spatial step $\Delta\theta = (L_1 + L_2)/(N - 1)$, and  $X^n = n\Delta X$, where $N$ is the total number of grid points in the interval and $\Delta X$ is the time increment.
Respectively, $\eta_i^n$ and $\eta_i^{n+1}$ denote the value of $\eta$ at the \textit{i}-th spatial point and ''time'' stages $X^n$ and $X^{n+1}$ correspondingly. We construct the following nonlinear difference scheme which is convergent:
\begin{multline}\label{eq:MS1}
\frac{\tilde{A}^n}{\Delta X}(\eta_i^{n+1}-\eta_i^n) + \frac{\tilde{B}^n}{2}(\eta_i^{n+1}+\eta_i^n)\\
+\frac{\tilde{C}^n}{4(\Delta\theta)^3} (-\eta_{i-2}^{n+1}+2 \eta_{i-1}^{n+1}-2 \eta_{i+1}^{n+1} + \eta_{i+2}^{n+1} -\eta_{i-2}^n+2\eta_{i-1}^n-2\eta_{i+1}^n + \eta_{i+2}^n)\\
+\frac{\tilde{D}^n}{8(\Delta\theta)}\big((\eta_{i+1}^{n+1})^2-(\eta_{i-1}^{n+1})^2 + (\eta_{i+1}^n)^2-(\eta_{i-1}^n)^2\big)=0.
\end{multline}

Such a scheme is stable when time-stepping with respect to the physical time, provided the iterative procedure to resolve the nonlinear terms is convergent. Since the scheme given by (\ref{eq:MS1}) cannot be implemented directly, because it is
nonlinear we follow the idea of \cite{CDM} to introduce internal iterations, namely
\begin{multline}\label{eq:MS11}
\frac{\tilde{A}^n}{\Delta X}(\eta_i^{n+1,k+1}-\eta_i^n) + \frac{\tilde{B}^n}{2}(\eta_i^{n+1,k+1}+\eta_i^n)\\
+\frac{\tilde{C}^n}{4(\Delta\theta)^3} (-\eta_{i-2}^{n+1,k+1}+2 \eta_{i-1}^{n+1,k+1}-2 \eta_{i+1}^{n+1,k+1} + \eta_{i+2}^{n+1,k+1} -\eta_{i-2}^n+2\eta_{i-1}^n-2\eta_{i+1}^n + \eta_{i+2}^n)\\
+\frac{\tilde{D}^n}{8(\Delta\theta)}\big(\eta_{i+1}^{n+1,k+1}\eta_{i+1}^{n+1,k}-\eta_{i-1}^{n+1,k+1}\eta_{i-1}^{n+1,k} + (\eta_{i+1}^n)^2-(\eta_{i-1}^n)^2\big)=0.
\end{multline}

This way, for the current iteration of the unknown function (superscript $n + 1; k +1$) we have an implicit system with five-diagonal band matrix. We begin from an initial condition $\eta^{n+1,0}=\eta^n$ and conduct the internal iterations (repeating the calculations for the same time step $(n + 1)$ with increasing value of the superscript $k$) until convergence.

Note that the initial condition is a very good guess which is within $\mathcal{O}(\Delta X)$ of the sought solution for $\eta$. This makes the convergence of the internal iterations very fast. We have performed the numerical experiments to verify this fact. Even
for very large values of the time increment $\Delta X$ we have not encountered any instability of the internal iterations. In each case under consideration we selected $\Delta X$ such that no more than six internal iterations were required to reach the precision of $10^{-12}$.
After the internal iterations converge, one gets the solution of the nonlinear scheme by setting $\eta^{n+1} \equiv \eta^{n+1,k+1}$. In such a way we have fully implicit, nonlinear and conservative scheme. For the inversion of the five-diagonal $N\times N$ matrix we use a generalized algorithm based on Gaussian elimination with pivoting (for details see \cite{read, TodChri13, marrak16}).

The scheme was thoroughly validated through the standard numerical tests involving halving the spacing and time increment. The global truncation error of time approximation was verified as by Runge principle and confirmed the second order of accuracy in time. In a similar fashion we found that the global spatial truncation error is also second-order, $\mathcal{O}((\Delta\theta)^2)$.

\end{document}